\documentclass[draft]{agujournal2019}
\usepackage{url} 
\usepackage{lineno}
\usepackage[inline]{trackchanges} 
\usepackage{soul}
\usepackage{amsmath, amssymb, amsfonts, stmaryrd}
\usepackage{multirow}
\draftfalse

\journalname{Journal of Advances in Modeling Earth Systems (JAMES)}

\begin{document}

%
%


%

\title{Simulating Atmospheric Processes in Earth System Models and Quantifying Uncertainties with Deep Learning Multi-Member and Stochastic Parameterizations}

%
%




\authors{Gunnar Behrens \affil{1,2}, Tom Beucler \affil{3,4},  Fernando Iglesias-Suarez \affil{1,5}, Sungduk Yu \affil{6,7}, Pierre Gentine \affil{8,9}, Michael Pritchard \affil{6,10}, Mierk Schwabe\affil{1}, Veronika Eyring \affil{1,2}}

\affiliation{1}{Deutsches Zentrum für Luft- und Raumfahrt (DLR), Institut für Physik der Atmosphäre, Oberpfaffenhofen, Germany}
\affiliation{2}{University of Bremen, Institute of Environmental Physics (IUP), Bremen, Germany}
\affiliation{3}{Faculty of Geosciences and Environment, University of Lausanne, Switzerland}
\affiliation{4}{Expertise Center for Climate Extremes, University of Lausanne, Switzerland}
\affiliation{5}{Predictia Intelligent Data Solutions S.L., Santander, Spain}
\affiliation{6}{Department of Earth System Science, University of California Irvine, Irvine, CA, USA}
\affiliation{7}{Multimodal Cognitive AI Research, Intel Labs, Santa Clara, CA, USA}
\affiliation{8}{Department of Earth and Environmental Engineering, Columbia University, New York, NY 10027, USA}
\affiliation{9}{Earth Institute and Data Science Institute, Columbia University, New York, NY 10027, USA}
\affiliation{10}{NVIDIA, USA}





\correspondingauthor{Gunnar Behrens}{gunnar.behrens@dlr.de}



\begin{keypoints}
    \item For deep learning, subgrid parameterizations, multi-member methods improve uncertainty quantification compared to dropout
    
    \item Offline multi-member parameterizations improve convective processes in the planetary boundary layer compared to counterparts 
    
    \item Online multi-member parameterizations allow stable hybrid simulations for 5 months, with temperature biases but better extreme precipitation   
\end{keypoints}

%
%

%
%


\begin{abstract}

Deep learning is a powerful tool to represent subgrid processes in climate models, but many application cases have so far used idealized settings and deterministic approaches. Here, we develop stochastic parameterizations with calibrated uncertainty quantification to learn subgrid convective and turbulent processes and surface radiative fluxes of a superparameterization embedded in an Earth System Model (ESM). We explore three methods to construct stochastic parameterizations: 1) a single Deep Neural Network (DNN) with Monte Carlo Dropout; 2) a multi-member parameterization; and 3) a Variational Encoder Decoder with latent space perturbation. We show that the multi-member parameterization improves the representation of convective processes, especially in the planetary boundary layer, compared to individual DNNs. The respective uncertainty quantification illustrates that methods 2) and 3) are advantageous compared to a dropout-based DNN parameterization regarding the spread of convective processes. Hybrid simulations with our best-performing multi-member parameterizations remained challenging and crash within the first days. Therefore, we develop a pragmatic partial coupling strategy relying on the superparameterization for condensate emulation. 
Partial coupling reduces the computational efficiency of hybrid Earth-like simulations but enables model stability over 5 months with our multi-member parameterizations. However, our hybrid simulations exhibit biases in thermodynamic fields and differences in precipitation patterns. Despite this, the multi-member parameterizations enable improvements in reproducing tropical extreme precipitation compared to a traditional convection parameterization. Despite these challenges, our results indicate the potential of a new generation of multi-member machine learning parameterizations leveraging uncertainty quantification to improve the representation of stochasticity of subgrid effects.

\end{abstract}

\section*{Plain Language Summary}

New artificial intelligence (AI)-algorithms that actively learn the influence of clouds on weather and climate have outperformed the skill of traditional schemes in recent years. However, instead of describing the complexity of cloud processes with their fine-scale variability (which is called \textit{stochasticity} in climate science), most of these AI algorithms only output a single deterministic prediction. This leads to reduced performance when stochasticity plays a large role. Here, we improve such AI algorithms so that they can generate multiple predictions (an \textit{ensemble}) from only one set of large-scale environmental conditions. When combined, the ensemble mean of this ensemble outperforms individual ones for variables like moisture on a hold-out dataset. The ensemble also allows us to quantify uncertainty. When we initially try to couple our ensemble methods back to the host climate model, the climate model crashes. To avoid these early model crashes, we implemented a costly work-around. This strategy stabilizes the climate model over 5 months. We find biases in temperature, moisture, and precipitation fields with our ensemble methods. However, heavy precipitation in the tropics is improved with our ensemble methods. Despite limitations, our methods show promising pathways for quantifiable uncertainty and stochastic approaches in data-driven parameterization methods. 


%
%

\section{Introduction}

Earth System Models (ESMs) are the main tools to project climate change. Despite notable improvements in simulating the climate of the recent past in the Coupled Model Intercomparison Project Phase 6 (CMIP6, \citeA{Eyring2016}), longstanding biases of convective processes still exist, such as the double Intertropical Convergence Zone (ITCZ) bias \cite{Bock2020,Lauer2023}. This demonstrates limitations in both our understanding of and our ability to simulate the Earth system. These limitations arise mainly from the current representation of subgrid convective processes via parameterizations in ESMs \cite{Gentine2021}. These parameterizations, traditionally empirical approximations of the subgrid mean effect on the state of the system, are necessary as the majority of convective processes occur on typical length scales much smaller than the standard horizontal grid size of an ESM ($\sim$ 100~km). Storm Resolving Models (SRMs, \citeA{Stevens2019}) partially overcome this ``convection parameterization deadlock'' \cite{Gentine2018} as they can resolve deep convection on their grid sizes of $\sim$ 2 - 10 km. By resolving parts of deep convective cells, SRMs improve the simulation of tropical precipitation \cite{Stevens2020}. These simulations are nonetheless computationally demanding, limiting model runs to no more than a few years \cite{Hohenegger2023}. Also, SRMs still rely on parameterizations for many important climate processes e.g., for shallow convection and small-scale turbulence in the planetary boundary layer. These fine-scale processes are critical for climate sensitivity \cite{Schneider2017}.

In an effort to overcome computational limits while retaining the quality of the representation of deep convective processes in SRMs, machine learning approaches have been developed to replace existing convection schemes in coarse-resolution ESMs \cite{Gentine2021,Eyring2024}. A superparameterization (SP) consists of nested high-resolution columns directly simulating subgrid turbulence, convection and convection-related radiative processes \cite{Grabowski2001,Khairoutdinov2001}. Emulating with machine learning such an SP in the Community Atmosphere Model in an aquaplanet setup was a trailblazing example in Earth system science in recent years \cite{Gentine2018}. This initial study was followed by several other studies coupling deep-learning subgrid parameterizations within general circulation models \cite{Rasp2018,Brenowitz2019,Yuval2020}, showing the potential of retaining many features of SRMs or an SP. Despite these advances it has to be noted that deep learned parameterizations heavily depend on the data sets they are trained on. For example, machine learning parameterizations trained on a superparameterized high-resolution simulation \cite{Khairoutdinov2001} might reproduce the known pronounced double ITCZ bias of the SP \cite{Woefle2018}. Machine learning subgrid parameterizations are also capable to represent convective processes from global SRM simulations or an SP with realistic topography \cite{Han2020,Mooers2021,Grundner2022,Han2023,Clark2022,Kwa2023,Watt_Meyer_2024}. Such realistic deep learning subgrid parameterizations can also at times be stable when coupled back to the coarse general circulation model, enabling decade-long prognostic model runs \cite{Wang2022}. Furthermore, a member of an ensemble deep learned emulator of the SP ran stably in the Community Atmosphere Model coupled to the Community Land Model over a few years with realistic boundary forcing \cite{Han2023}. A different avenue of deep learning physical tendencies based on reanalysis data recently enabled stable neural circulation model integration over 40 years on coarse climate model time scales \cite{Kochkov2024}. These recent advances constitute a step on the long way towards operational machine learning parameterizations in ESMs.

Despite these advances, several caveats remain concerning the use of machine learning-based subgrid parameterizations especially in the planetary boundary layer and adjacent layers of the lower troposphere \cite{Gentine2018,Mooers2021,Behrens2022}. It has been hypothesized that this relatively low reproduction skill may be associated with limitations of \textit{deterministic} deep learning algorithms, i.e., failing to capture turbulent and stochastic features of convective processes in the lower troposphere (e.g. \citeA{Mooers2021,Behrens2022}). Deterministic parameterizations implicitly assume that the resolved state of the system completely determines the effects related to the unresolved processes. However, in reality different subgrid convective effects may result from the same environmental conditions \cite{Franzke2014,Christensen_2024} because of internal stochasticity. 
Therefore, it is natural to wonder whether ensemble-based deterministic predictions \cite{Jones2019a,Jones2019b} and stochastic prediction approaches \cite{Berner2017,Palmer2019,Christensen_2024} can better capture both the mean effect and the stochastic nature of convective processes. While an ensemble-based deterministic approach involves a set of deterministic predictions (mean effect, \citeA{Jones2019a}), there are different stochastic prediction approaches, e.g., random deterministic predictions subsampling or added noise, to introduce a chaotic behavior into the predictions \cite{Berner2017,Palmer2019}. Ensemble-based deterministic predictions may improve the simulated climate  mean state, for instance by improving the representation of precipitation averages compared to observations, but at the same time may decrease the simulated spatio-temporal variability \cite{Jones2019b}.      

Stochastic approaches, such as stochastic perturbed parameter ensembles, have improved weather forecast skill and spread, and are beneficial for uncertainty quantification and data assimilation \cite{Christensen2015}. These techniques have also started to be applied in climate science \cite{Haynes2023,Christensen_2024}, ranging from idealized to more realistic frameworks, and have been shown to reduce model bias and to better represent long-term climate variability \cite{Berner2017}. Several studies focusing on the Lorenz 96 model showed the superiority of stochastic parameterizations over their deterministic counterparts \cite{Gagne2020,Parthipan2022,Bhouri2022}. A Monte Carlo Dropout stochastic machine learning entrainment and detrainment scheme for shallow convective processes outperformed traditional schemes \cite{Shin2022}. A Conditional Generative Adversarial Network reproduced the spread and general statistics of the heating and moistening profiles due to convection with high accuracy over the tropical Pacific \cite{Nadiga2022}. A stochastic multi-plume mass-flux parameterization of dry and shallow convection improved the representation of shallow cumulus convection \cite{Chinita2023}. Combining a deterministic mass flux closure with stochastic sampling of the cloud base mass fluxes corrected the spatial and temporal distribution of cloudiness in an SRM model run \cite{Sakradzija2018}. In ocean modeling, stochastic machine learning approaches captured the effects of mesoscale eddies with high skill, enabling more realistic energy cascades from the large-scale towards the mesoscale on ocean grid resolutions coarser than the Rossby radius of deformation \cite{Guillaumin2021,Perezhogin2023}. These examples indicate the potential of stochastic approaches to advance state-of-the-art parameterizations in ESMs.
In a similar fashion \citeA{Kochkov2024} showed that stochastic deep learning based forecasts have the potential to outperform numerical weather prediction models for certain variable fields.

In this study, we develop a new approach to test the potential of stochastic and multi-member deep learning subgrid parameterizations of convection in a superparameterized ESM with a realistic configuration. We calibrate the stochasticity by evaluating uncertainty quantification (i.e. using the predictions of the different ensemble members) of the developed schemes for the multi-variate output data set of the SP. Offline (evaluated against test data), our results show that the multi-member predictions, both deterministic and stochastic, improve the representation of convective processes compared to individual deterministic deep learning members within the planetary boundary layer. With regard to the stochastic parameterizations, we demonstrate that both a multi-member parameterization and the perturbation of a Variational Encoder Decoder's (VED) latent space provide better uncertainty spread compared to traditional Monte Carlo dropout methods, in line with the findings of \citeA{Haynes2023}. Online (when the parameterization is coupled to the ESM), our deterministic and stochastic multi-member parameterizations introduce biases in the thermodynamic fields of the coarse ESM but showed some improvements of precipitation biases compared to its counterpart with traditional parameterizations.

The manuscript is structured as follows. Section \ref{sec:climate_model} describes our climate modeling setup. Section \ref{sec:Deep_Learning_approaches} covers our deep learning parameterizations, including the description of the deterministic and stochastic approaches. Section \ref{sec:Offline_Results} provides a detailed evaluation of our parameterizations \textit{before} coupling them to the host ESM, referred to as ``offline'' evaluation. Section \ref{sec:online} assesses the performance of our parameterizations \textit{after} their integration within the host ESM (referred to as ``online'' performance) and addresses related caveats. Section \ref{sec:Conclusion} provides a summary and discussion of the added value of our deep-learned deterministic and stochastic multi-member parameterizations in the broader context of Earth system modeling.

\section {Climate Modeling Setup \label{sec:climate_model}}

In this study we use the Super-Parameterized Community Earth System Model Version 2.1.3 (SPCESM2, \citeA{Danabasoglu2020}) for the construction of our stochastic and deterministic parameterizations. The atmospheric component of CESM2 is the Community Atmosphere Model version 6 (CAM6). In our configuration CAM6 is run without interactive chemistry, and thus radiatively-active aerosols and gases are prescribed. CAM6 has a horizontal grid size of $2.5^{\circ} $ $\times $ $ 1.875^{\circ}$ ($144 \times 96$ grid cells). The vertical axis consists of 26 levels on a hybrid sigma-pressure grid with 14 tropospheric levels (p $>$ 200 hPa). CAM6 has a timestep of 1800 s. To represent subgrid processes (convection, subgrid radiative effects, and fine-scale eddies) in each grid cell of CAM6, we use an SP \cite{Khairoutdinov2001,Grabowski2001}. SP, also known as multiscale modeling framework (MMF, i.e. \citeA{Yu2023_neurips}), consists of 32 nested two-dimensional grid columns with a finer horizontal resolution of 4 km, which partially resolves deep convection and associated gravity waves. These grid columns are meridionally oriented (north to south) as described in \citeA{Pritchard2014b}. SP and CAM6 share the same vertical discretization after an initial interpolation at the beginning of each SP time step (20 s), from the 24 levels of SP to the CAM6 vertical axis. 
Our configuration of SP uses a Smagorinsky 1.5-order turbulence scheme to parameterize fine-scale turbulence and a one-moment microphysics scheme \cite{Khairoutdinov2001,Grabowski2001}. The microphysics scheme allows the separation into cloud ice and liquid water phase and respective phase tendencies. Horizontal advection of high-resolution convection-related fields (momentum, cloud condensates) from the nested SP to the neighbouring CAM6 cells' nested SP is neglected. Instead the advection of these convection-related fields is handled via the dynamical core of the coarse CAM6 model with known limitations \cite{Jansson2022}.

The atmosphere is coupled to the land component (Community Land Model version 5, CLM5), which includes realistic topographic boundary conditions. We use prescribed sea surface temperatures and sea ice fields (Merged Hadley-NOAA/OI Sea Surface Temperature and Sea-Ice Concentration, \citeA{Hurrell2008}). Our simulations are driven by observed solar spectral irradiance and concentrations of aerosols and atmospheric trace gases (e.g., ozone). For a more detailed description of CESM2, we point the interested reader to \citeA{Danabasoglu2020}, and for SP to \citeA{Khairoutdinov2001}. The SPCESM2 version used here can be found on GitHub (\url{https://github.com/SciPritchardLab/CESM2-ML-coupler}).

The next section explains the deep learning approaches we developed to build a stochastic or a multi-member, data-driven emulator of SP.

\section{Deep Learning Parameterizations \label{sec:Deep_Learning_approaches}}
In this section, we first describe the general approach to the training of the deep learning subgrid processes in SPCESM2 (Sec~\ref{sec:Gen_appr}). We then describe the deep learning (DL) algorithms (Sec~\ref{sub:deep_learning_algo}), before constructing and calibrating stochastic and deterministic DL parameterizations (Sec~\ref{sub:stochastic_param}). 
Table \ref{tab:overview_stoch_deter} gives an overview of our developed stochastic and deterministic parameterizations. Moreover it helps the reader understand the acronyms of the different models that we will use in the following.   

\subsection{Problem Statement \label{sec:Gen_appr}}

 Our DL parameterizations aim to represent the \textit{aggregate} effect of subgrid processes, as simulated by the SP component of SPCESM. To achieve this, our DL algorithms predict a grid-averaged subset of SP's subgrid variables based on the large-scale atmospheric conditions modeled by CAM6, hereafter referred to as ``CAM variables''. During the DL-coupled climate model simulations, these predicted subgrid variables (i.e., vertical profiles of subgrid specific humidity and temperature) are used to couple the atmospheric model with the other components at the surface (e.g., CLM5 land model and boundary conditions from the ocean model). 

\begin{figure}[t]
    \centering
    \includegraphics[width=\textwidth]{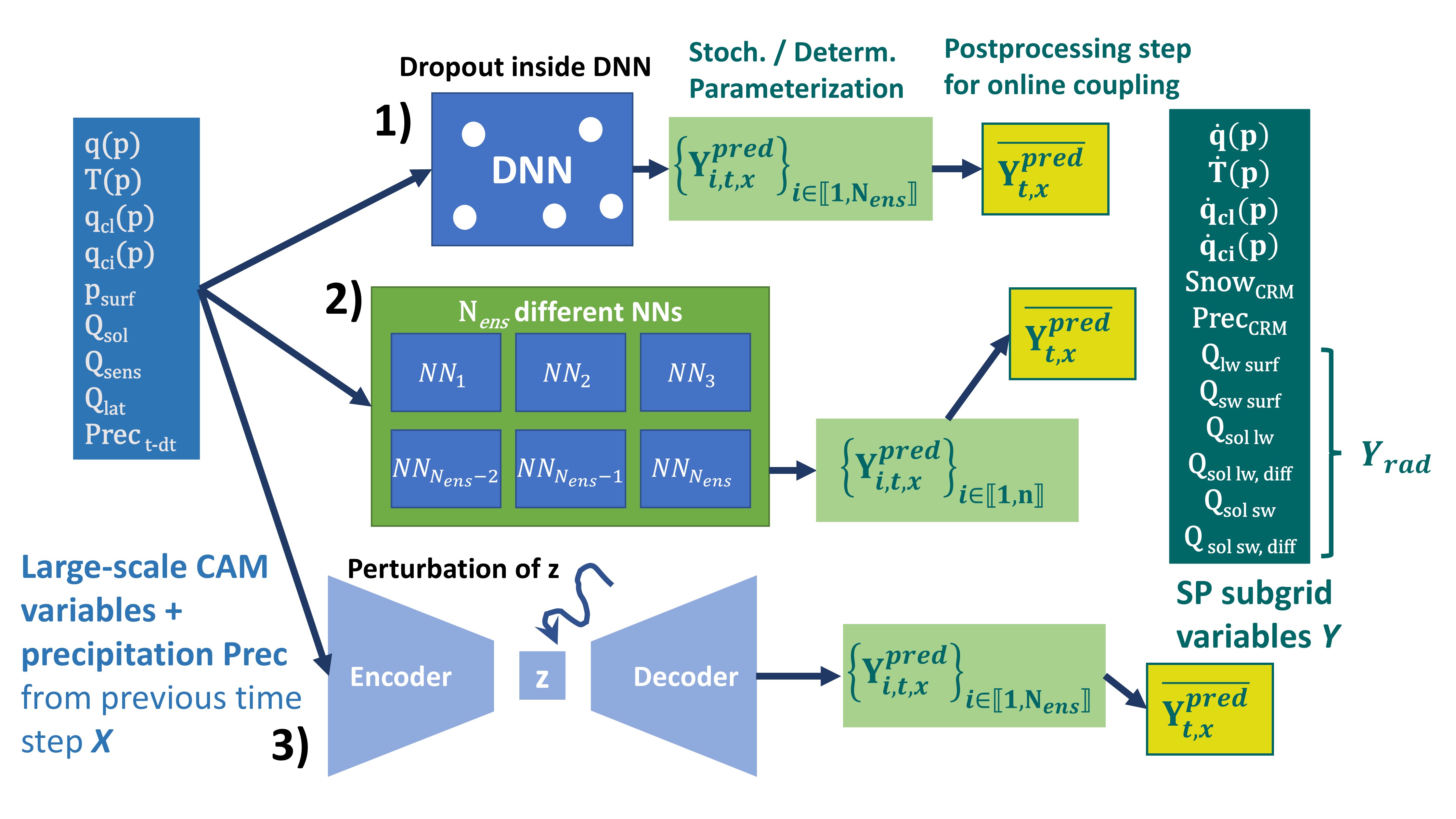}
    \caption{Architectures of the stochastic parameterization strategies for reproducing the superparameterization: We compare three stochastic parameterization strategies for reproducing the superparameterization (SP), which simulates SP subgrid variables ($\boldsymbol{Y}$) based on the large-scale Community Atmosphere Model (CAM) variables ($\boldsymbol{X}$): 1) Applying Monte-Carlo dropout to a single deep neural network (DNN) to generate a prediction based on the mean of $\mathrm{N}_{ens}$ draws. 2) Employing a subset of $\mathrm{n}$ neural networks randomly drawn from a set of $\mathrm{N}_{ens}$ deterministic neural networks to generate $\mathrm{n}$ predictions that can be averaged for the final prediction. 3) Perturbing the latent space of a Variational Encoder-Decoder network $\mathrm{N}_{ens}$ times to produce $\mathrm{N}_{ens}$ predictions that are subsequently averaged. In addition, Table \ref{tab:overview_stoch_deter} gives an overview of our developed deep learning ensemble parameterizations.}
    \label{fig:intro_schematic}
\end{figure}

The input data closely follows the CAM standard large-scale variables except for one additional variable, the previous time step's precipitation, ($\mathrm{Prec}_{t-dt}$) which was helpful for the performance of the DL algorithms. The input $\boldsymbol{X}$ (Fig. \ref{fig:intro_schematic}) is a stacked vector of size 109 and is given by: 
\begin{equation}
\boldsymbol{X}=\begin{bmatrix}\boldsymbol{q}\left(\boldsymbol{p}\right) & \boldsymbol{T}\left(\boldsymbol{p}\right) & \boldsymbol{q_{cl}}\left(\boldsymbol{p}\right) & \boldsymbol{q_{ci}}\left(\boldsymbol{p}\right) & p_{\mathrm{surf}} & Q_{\mathrm{sol}} & Q_{\mathrm{sens}} & Q_{\mathrm{lat}} & \mathrm{Prec}_{t-dt}\end{bmatrix}^{T},
\end{equation}
where $\boldsymbol{X} $ includes the 4 vertical profiles (with 26 vertical levels) of specific humidity $\boldsymbol{q}\left(\boldsymbol{p}\right) $ [g/kg], temperature $\boldsymbol{T}\left(\boldsymbol{p}\right) $ [K], cloud liquid water content $\boldsymbol{q_{cl}}\left(\boldsymbol{p}\right) $ [g/kg], and cloud ice water content $\boldsymbol{q_{ci}}\left(\boldsymbol{p}\right) $ [g/kg]. Additionally, $\boldsymbol{X}$ comprises the scalar values of surface pressure $p_{\mathrm{surf}} $ [hPa], solar insolation Q$_{sol}$ [$\mathrm{W}/\mathrm{m}^2 $], surface sensible $Q_{\mathrm{sens}} $ [$\mathrm{W}/\mathrm{m}^2 $] and latent heat flux $Q_{\mathrm{lat}} $ [$\mathrm{W}/\mathrm{m}^2 $] from the current timestep. Additionally we use the previous timestep's precipitation $\mathrm{Prec}_{t-dt} $ [mm/h] as input to complement the other CAM variables. Including $\mathrm{Prec}_{t-dt} $ strongly improves the prediction of near-surface heating and moistening tendencies that are of great importance for the coupling to the CLM5 land model, which is aligned with the findings of previous studies \cite{Han2020,Han2023}. 

\begin{table}
    \begin{minipage}{\textwidth}
    \centering
    \bgroup\renewcommand{\arraystretch}{1.3}
    \begin{tabular}{p{0.3cm}|p{2.5cm}|p{1.9cm}|p{1.1cm}|p{3.3cm}|p{3.0cm}}
    \textbf{} &  \textbf{Acronym \mbox{climate} model}& \textbf{Acronym parameterization} & \textbf{No. Networks} &\textbf{Method}& \textbf{Stochastic \mbox{parameter}} \\ \hline 
    \rotatebox[origin=r]{90}{\parbox{20mm} {\multirow{2}{0.5em}{\textbf{deterministic}}}} &$\overline{\mathrm{DNN}}$-SP-CESM\footnote{for this CESM run SP's predictions of cloud condensate tendencies instead of $\overline{\mathrm{DNN}}$'s predictions are used. } & $\overline{\mathrm{DNN}}$ & 7 & deterministic ensemble mean prediction & - \\ 
    \cline{2-6}
    & - & $\overline{\mathrm{VED}}$ & 6 & deterministic ensemble mean prediction & - \\
    \hline
    \rotatebox[origin=r]{90}{\parbox{15mm}{\multirow{1}{0.5em}[-1em]{\textbf{stochastic}}}} &  - & DNN-dropout\footnote{dropout including 7 samples per prediction of DNN 1 (Supporting Information Tab. S3)} & 1 & dropout & dr=0.01 \\ \cline{2-6}
    & DNN-ens-SP-CESM\footnote{for this CESM run SP's predictions of cloud condensate tendencies instead of DNN-ensemble's predictions are used.} & DNN-ensemble\footnote{based on all DNNs (Tab. S3), the 5 out of 7 members are randomly drawn for each timestep and location} & 7 & ensemble & randomly draw 5 out of 7 members for averaging \\ \cline{2-6}
    & - & VED-draws\footnote{based on 7 predictions of VED 1 (Tab. S4)} & 1 & latent space reparameterization & 7 draws \\ \cline{2-6}
    & - & VED-static\footnote{based on 7 predictions of VED 1 (Tab. S4) with scalar $\alpha=0.5$} & 1 & latent space perturbation & 7 draws with scalar $\alpha=0.5$\\ \cline{2-6}
    & - & VED-varying\footnote{based on 7 predictions of VED 1 (Tab. S4) with $\alpha$-array} & 1 & latent space perturbation & 7 draws with $\alpha$-array\\ 
    \hline
    \rotatebox[origin=r]{90}{\parbox{13mm}{\multirow{1}{0.5em}[-1em]{\textbf{reference}}}}& SP-CESM& - & - & superparameterization & \\\cline{2-6}
    & ZM-CESM & - &  & Zhang-McFarlane scheme\footnote{For this run the Zhang-McFarlane convection scheme \cite{ZhangMcFarlane1995} is used.} & \\
    \end{tabular}
    \egroup
    \end{minipage}
    \caption{Summary of the deterministic (two top rows) and stochastic parameterizations (third - seventh row) we developed, and reference schemes (bottom two rows) we used. The 1$^{st}$ column shows the split into deterministic, stochastic and reference schemes. The 2$^{nd}$ and 3$^{rd}$ column indicate the acronyms of the respective parameterizations in the Community Earth System Model (CESM; section \ref{sec:online}) and in our offline evaluation (section \ref{sec:Offline_Results}). The other columns show for each parameterization the number of DL networks used, the method used to generate the predictions, and key stochastic parameters for the stochastic parameterizations. This table is accompanied by Table S7 that shows observational products used for the evaluation of our CESM runs.}
    \label{tab:overview_stoch_deter}
\end{table}

The output vector ($\boldsymbol{Y}$, predictants or target) of our data-driven parameterization has a length of 112 (Fig.~\ref{fig:intro_schematic}) and is given by:
\begin{equation}
    \boldsymbol{Y}=\begin{bmatrix}\boldsymbol{\dot{q}}\left(\boldsymbol{p}\right) & \boldsymbol{\dot{T}}\left(\boldsymbol{p}\right) & \boldsymbol{\dot{q}_{cl}}\left(\boldsymbol{p}\right) & \boldsymbol{\dot{q}_{ci}}\left(\boldsymbol{p}\right) & \mathrm{Snow}_{\mathrm{CRM}} &  \mathrm{Prec}_{\mathrm{CRM}} & \boldsymbol{Y_{\mathrm{rad}}}\end{bmatrix}^{T},
\end{equation}
where $\boldsymbol{Y} $ includes the 4 vertical profiles of: specific humidity tendency $\boldsymbol{\dot{q}}\left(\boldsymbol{p}\right) $ [$\frac{g}{kg \times h}$], temperature tendency $\boldsymbol{\dot{T}}\left(\boldsymbol{p}\right) $ [K/h], cloud liquid water tendency $\boldsymbol{\dot{q}_{cl}}\left(\boldsymbol{p}\right) $ [$\frac{g}{kg \times h}$], and cloud ice water tendency $\boldsymbol{\dot{q}_{ci}}\left(\boldsymbol{p}\right) $ [$\frac{g}{kg \times h}$]. Here, we use ``tendency'' and the notation $\dot{y} $ as a shorthand for the difference between the values of state variables before and after the SP call, normalized by the CAM6 time step ($dt= 1800 $\,s, see e.g., SI equation~S1). Note that this call precedes and does not include the calculations for surface coupling. $\boldsymbol{Y} $ further includes the cloud-resolving precipitation ($\mathrm{Prec}_{\mathrm{CRM}} $) and snow rates ($\mathrm{Snow}_{\mathrm{CRM}} $), both simulated by SP and expressed in units mm/h. To facilitate reading, we grouped all radiative outputs required for coupling to the surface in $ \boldsymbol{Y_{\mathrm{rad}}}  $:
\begin{equation}
\label{eq:Y_rad}
    \boldsymbol{Y_{\mathrm{rad}}}=\begin{bmatrix}Q_{\mathrm{lw \ surf}} & Q_{\mathrm{sw \ surf}}& Q_{\mathrm{sol \ lw}} & Q_{\mathrm{sol \ lw,\ diff}} & Q_{\mathrm{sol \ sw}} & Q_{\mathrm{sol \ sw,\ diff}}\end{bmatrix}^{T},
\end{equation}
where $ \boldsymbol{Y_{\mathrm{rad}}}$ includes the downward surface longwave heat flux $Q_{\mathrm{lw \ surf}} $, the downward surface shortwave heat flux $Q_{\mathrm{sw \ surf}}$, the near-infrared part of the downward solar radiation at the surface, decomposed into its direct ($Q_{\mathrm{sol \ lw}} $) and diffuse ($Q_{\mathrm{sol \ lw,\ diff}} $) components, and the direct ($Q_{\mathrm{sol \ sw}} $) and diffuse ($Q_{\mathrm{sol \ sw ,\ diff}} $) components of the solar radiation's visible wavelengths; all are in units of $\mathrm{W}/\mathrm{m}^2$. In the following, we couple the predictions of the surface radiative fluxes $\boldsymbol{Y_{\mathrm{rad}}}$ to CESM2 to investigate also the stability of CESM2 with such deep learned radiative fluxes. This contrasts our work to \citeA{Han2023} that sidestepped the coupling of these crucial terms. Apart from these radiative fluxes, momentum and mass fluxes are also used to couple CAM6 to the surface.  

For DL algorithms that involve multiple input and output variables with different physical units, a suitable normalization is important for both inputs ($\boldsymbol{X}$) and outputs ($\boldsymbol{Y}$), as normalization choices affect the variables' relative importance during training. We normalize each of the inputs by subtracting its mean and dividing the resulting difference by the corresponding range, resulting in normalized inputs between -1 and 1. We normalize each output variable using a reference standard deviation as in \citeA{Behrens2022} (see SI section 1.2 for details).

To avoid spatiotemporal correlations and overfitting \cite{Behrens2022,Rasp2018}, we extract 84 days per year, specifically 7 consecutive days from each month, for training (Year 2013) shuffled in space and time, validation (2014), and testing (2015). These data originate from a historical SPCESM2 run spanning 2003 to 2015, ensuring the exclusion of any model spin-up effects. Each dataset contains 55,572,480 samples, and is balanced with respect to the diurnal and seasonal cycles. We test also a configuration without shuffling to introduce some spatio-temporal correlation, but this resulted in generally weaker reproduction skills and less skillful uncertainty quantification.

\subsection{Machine Learning Algorithms \label{sub:deep_learning_algo}}

To map $\boldsymbol{X}$ to $\boldsymbol{Y}$, we implement two different model types ({Tab.~1}): Deep Neural Networks (DNNs) and Variational Encoder Decoder (VED, \citeA{Kingma2014}) structures, which use a lower-dimensional manifold between the encoding and decoding part of the network, also known as ``latent space'' in data science. In the following we will briefly describe these two network types and the associated hyperparameter searches we conducted. We will use these two neural network types to build stochastic and deterministic DL parameterizations (see section \ref{sub:stochastic_param}). 

\subsubsection{Deep Neural Networks}

DNNs consist of a set of fully connected layers that contain nodes. These nodes perform a non-linear regression task, and their weights and biases are optimized to reduce a loss function. The linear sum of the previous layer is then passed through a non-linear function, referred to as an ``activation function''. Our DNNs have an input layer of 109 nodes ($\boldsymbol{X}$'s length) and an output layer of $\mathrm{N}_{outputs} = 112$ nodes ($\boldsymbol{Y}$'s length). To optimize the DNNs' weights and biases, we use the mean-squared error (MSE) between the predictions (\textbf{Y$^{pred}$}) and the original data ($\boldsymbol{Y}$) as our loss function (Equation~\ref{eq:MSE}):

\begin{linenomath*}
 \begin{equation}
\mathrm{MSE}\left(\boldsymbol{Y},\boldsymbol{Y^{\mathrm{pred}}}\right)=\frac{1}{N_{\mathrm{outputs}}\times N_{\mathrm{batches}}}\sum_{k=1}^{N_{\mathrm{batch}}}\sum_{j=1}^{N_{\mathrm{outputs}}}\left(Y_{j,k}-Y_{j,k}^{\mathrm{pred}}\right)^{2},
\label{eq:MSE}
 \end{equation}
 \end{linenomath*}
where $\mathrm{N}_{batch}$ is the batch size (i.e., the number of samples fed to the network per backpropagation step), $Y_{j,k}^{\mathrm{pred}}$ is the network's prediction of the j-th output for the k-th sample in the batch, and $Y_{j,k} $ the corresponding target value we aim to predict. 

To optimize the overall setup of the DNNs we conducted an extensive hyperparameter search, including the batch size, the learning rate (i.e., the down-gradient step with respect to the loss function for the network optimization during training), the number of nodes per layer (integral parts of the network, which determines the number of weights and biases to be optimized during training), the number of hidden layers (network layers between the input and output layer), and the activation function (see SI section~S.1.1). We find that the performance of DNNs is most sensitive to changes in learning rate and batch size. Other predefined settings of our DNNs are the use of Adam (\citeA{Kingma2014Adam}) as optimizer (an algorithm that improves the network performance during training) and a predefined learning rate schedule (which decreases the initial learning rate after a certain epoch, see {SI section~S.1.1 for details}). The hyperparameters of the 7 best-performing DNNs are summarized in {SI Tab.~S3}. We will use these DNNs as building blocks of our deterministic and stochastic parameterizations and compare them to VEDs, which we describe next.

\subsubsection{Variational Encoder Decoders}
\label{subsubsec:VED_theory}

Similar to DNNs, the VEDs comprise input and output layers and dense fully connected hidden layers. The main difference between the VEDs and DNNs is the dimensionality reduction within the network into a small latent space and the addition of stochasticity in the latent space. The encoding part (Encoder) compresses the information down to the latent space through hidden layers with successively smaller node numbers from layer to layer (see Fig. \ref{fig:intro_schematic}). This latent space is a lower-order representation of the original information with a latent space width of size $\mathrm{N}_{latent}$, which is the number of nodes in the latent space. Within the latent space, the mean $\mu$ and logarithmic variance $\log \sigma^2$ of the latent distributions are optimized. $\mu$ and $\log \sigma^2$ are then mapped on an isotropic Gaussian, performing a ``reparameterization'', to generate the stochastic latent variables \textbf{z} \cite{Kingma2014,Behrens2022}. Different techniques can be used to interpret the encoded information with respect to the input and output data (see \citeA{Behrens2022,Mooers2022,Shamekh2023}). The latent variables \textbf{z} are then the input to the decoding part of the network (Decoder), which maps the information back to generate predictions. The VED's loss function ${\cal L}_{\mathrm{VED}} $ is the sum of the MSE loss function given by equation~\ref{eq:MSE} and a Kullback-Leibler (KL) loss term, which can be interpreted as a regularizer of the latent distribution towards a normal decorrelated distribution for disentanglement \cite{Kingma2014}:

\begin{equation}
    {\cal L}_{\mathrm{VED}}\left(\boldsymbol{Y},\boldsymbol{Y^{\mathrm{pred}}}\right)=\mathrm{MSE}\left(\boldsymbol{Y},\boldsymbol{Y^{\mathrm{pred}}}\right)+\lambda\times\underbrace{\frac{1}{2N_{\mathrm{batch}}}\sum_{k=1}^{N_{\mathrm{batch}}}\sum_{z=1}^{N_{\mathrm{latent}}}\left(\mu_{z,k}^{2}+\sigma_{z,k}^{2}-\ln\sigma_{z,k}^{2}-1\right)}_{\mathrm{KL\ Loss}},
\end{equation}
where the regularization factor (\textbf{$\lambda$}) regulates the weight given to the MSE and KL losses during training. We push this balance towards an enhanced reconstruction (smaller MSE) to the expense of the KL loss term (\textbf{$\lambda <1$}). In this study we use a static regularization factor, so a constant $\lambda$ that can be used as an additional hyperparameter. Our approach to construct the VED deviates from the standard data science approach of a probabilistic Variational (Auto)Encoder (VAE) decoder in two ways. First, we use the MSE (equation~\ref{eq:MSE}) between $\boldsymbol{Y^{pred}}$ and $\boldsymbol{Y}$ to measure the reconstruction error, instead of the squared error between predictions $\boldsymbol{X^{pred}}$ and true $\boldsymbol{X}$ input variables that is often used in the loss function of VAEs in data science (see e.g., \citeA{Mooers2022} for more details). This allows us to directly quantify how well the original convective processes in $\boldsymbol{Y}$ are reproduced. Secondly, the main focus of the training of our VED lies on an accurate reproduction and not on a perfect disentanglement inside the latent space of the VED, thus using a strong regularization of the KL loss. Such an enhanced determinism of the VEDs is beneficial to increase the general performance on the complex multi-input, multi-output data set of the superparamaterization compared to a fully probabilistic setup \cite{Yu2023_neurips}. 

The list of evaluated hyperparameters for the VEDs includes batch size, learning rate, number of nodes in the first or last hidden layer of the Encoder or Decoder, the number of hidden layers of the Encoder or Decoder, the latent space width and the regularization factor \textbf{$\lambda$}. We find that the VED's performance is most sensitive to the batch size, learning rate, latent space width and the regularization factor. Details about the conducted hyperparameter search and VED architecture can be found in the {SI (section~S.1.1 and Tab.~S4)}.

\subsection{Stochastic and Deterministic Ensemble Deep Learning Parameterizations \label{sub:stochastic_param}}

Here, we present three suitable approaches to develop a stochastic parameterization based on the machine learning algorithms introduced in the previous subsection (Fig. \ref{fig:intro_schematic}): dropout inside a DNN as a source of stochasticity, multi-member prediction of a number of neural networks, and a latent space perturbation of a single VED, inspired by the enhanced interpretability gained with latent space perturbations shown in \citeA{Behrens2022}.

\subsubsection{Dropout}

Dropout, also known as Monte Carlo Dropout (MCD), is widely applied to reduce overfitting, which is characterized by an elevated training performance compared to validation or test performance \cite{Hinton2012}. In addition, MCD can be used to quantify the uncertainty of predictions, and therefore to estimate stochasticity. It has been shown that the resulting uncertainty quantification and stochastic predictions of MCD have substantial limitations, in particular an underestimation of systematic spread and the inflation of deterministic errors compared to more complex methods to construct stochastic predictions \cite{Haynes2023}. 

With these caveats in mind, we use MCD as a simple baseline for our stochastic parameterizations. We apply MCD to one of the best-performing DNNs (DNN-dropout in Table \ref{tab:overview_stoch_deter} and hereafter) by adding a dropout layer after the last hidden layer of the network directly in front of the output layer. We choose a dropout rate $\boldsymbol{dr}$ of 0.01, meaning that 1\% of the input linkages to the  dropout layer are randomly discarded for each sample. While this small dropout rate underestimates the spread, higher values of the dropout rate (e.g., 0.05) significantly deteriorate reconstruction quality. We train the DNN with MCD and the hyperparameter settings of the DNN and use an active MCD also during testing. 

To construct an ensemble with MCD (Fig. \ref{fig:intro_schematic}) we repeat the sample-level prediction $\mathrm{N}_{ens}$ times (see equation~\ref{eq:Ensemble}), where $\mathrm{N}_{ens}$ is the ensemble size and $i$ symbolizes the i-th sampling of the DNN with active dropout $\boldsymbol{dr}$. Due to the active dropout the resulting ensemble $\{\boldsymbol{Y}^{pred}_{i,t,x}\}$ is of stochastic nature and provides uncertainty quantification for each timestep $t$ and grid cell $x$. We use the ensemble mean of the MCD ensemble (section \ref{susbsec:deter_perf}) and individual members (sections \ref{subsec:Uncert_quant}, \ref{subsec:CRPS}) to compare against other approaches to construct a stochastic and multi-member parameterization for CESM2.  

\begin{linenomath*}
\begin{equation}
\overline{\boldsymbol{Y}^{\mathrm{pred}}}=\frac{1}{N_{\mathrm{ens}}}\sum_{i=1}^{N_{\mathrm{ens}}}\boldsymbol{Y}_{i}^{\mathrm{pred}}\ \ ,\ \ \boldsymbol{Y}_{i}^{\mathrm{pred}}=\left(\mathrm{Best\ DNN}_{\mathrm{dr}}\right)_{i}\left(\boldsymbol{X}\right)
\label{eq:Ensemble}
\end{equation}
    
\end{linenomath*}

\subsubsection{Multi-Member Parameterizations}

Ensemble predictions are one common way to provide uncertainty quantification such as in weather forecasting \cite{Gneiting2005} or climate projections \cite{Eyring2016}, as climate and weather are governed by internal variability and stochasticity; some of them due to convective and turbulent processes \cite{Berner2017}. 
Inspired by these traditional climate modeling approaches, we develop multi-member stochastic and deterministic parameterizations using DNNs and VEDs (stochastic: DNN-ensemble; deterministic: $\mathrm{\overline{DNN}}$, $\mathrm{\overline{VED}}$ in Tab. \ref{tab:overview_stoch_deter}). To better differentiate between the various parameterizations, we denote this type of parameterizations as ``multi-member'' parameterizations throughout the manuscript. These  multi-member parameterizations will prove to have advantages relative to a single deterministic prediction of an individual neural network. In the following we use the terminology ``deterministic multi-member parameterization'' for a parameterization built without additional subsampling ($\mathrm{n} = \mathrm{N}_{ens}$ in equation 7, where $\mathrm{N}_{ens}$ is the maximum number of ensemble members and $\mathrm{n}$ is the used ensemble size).
To account for limitations when it comes to the computational overhead (see Tab. S6) and the applicability of the multi-member parameterizations, we restrict the ensemble size $\mathrm{n}$ to 7 members (A similar number of ensemble members as \citeA{Han2023}, who used an ensemble size of 8.). We note that this number of members is a critical hyperparameter for ensemble predictions, and larger (more diverse) ensembles yield often better performance over smaller ones with decreased spread between the ensemble members. Yet, larger ensembles have higher computational costs and  require larger memory so that they might not be practical (see Tab. S6). 

We generate either a deterministic ($\mathrm{n} = \mathrm{N}_{ens}$) or a stochastic multi-member parameterization ($\mathrm{n} < \mathrm{N}_{ens}$) (see equation 7) for each time step $t$ and grid cell $x$. In the stochastic case we randomly draw for each time step and grid cell a subset of members of size $\mathrm{n} < \mathrm{N}_{ens}$ out of the set used to generate the deterministic multi-member parameterization. Equation 7 shows the computation of the ensemble mean that we use for our online coupling experiments later on (Fig. \ref{fig:intro_schematic}),

\begin{linenomath*}
\begin{equation}
\overline{\boldsymbol{Y}^{\mathrm{pred}}}=\frac{1}{n}\sum_{i=1}^{n}{\boldsymbol{Y}_{i}^{\mathrm{pred}}}\ ,\ \ \boldsymbol{Y}_{i}^{\mathrm{pred}}=\mathrm{NN}_{i}\left(\boldsymbol{X}\right)
\end{equation}
\end{linenomath*}
where $\mathrm{n}$ elements are randomly drawn out of the multi-member parameterization $\mathrm{N}_{ens}$ in the stochastic case. Decreasing $\mathrm{n}$ towards 1 yields a larger degree of stochasticity. We tested the number of samples that are randomly drawn and found that 5 out of 7 members is a good compromise between added stochasticity and the overall reproduction skill of subgrid processes. For ensemble sizes smaller than 5 the general reproduction skill deteriorates. In the following we show the results of a DNN-based stochastic multi-member parameterization with 5 out of 7 members (DNN-ensemble, Tab. \ref{tab:overview_stoch_deter}), which illustrates the applicability of such an approach to generate stochasticity (Fig. \ref{fig:intro_schematic}). The added value of stochasticity for the offline performance is negligible based on the analysed offline metrics, but we see an improved reproduction of precipitation extremes with the DNN-ensemble in comparison to the deterministic $\mathrm{\overline{DNN}}$ multi-member parameterization when partially coupled to CESM2 later on.

\subsubsection{Latent Space Perturbation}
\label{subsec:lat_perturb}

This method is inspired by the interpretability and the potential of perturbing the latent space of the VED \cite{Behrens2022}.

We develop a two-step approach to build stochastic parameterizations via latent space perturbation. First, we train one of the best-performing VEDs ({Tab. S4}) to achieve a realistic reproduction of convection-related SP variables $\boldsymbol{Y}$. This particular VED is the base for the VED-static and VED-varying stochastic parameterizations (Tab. \ref{tab:overview_stoch_deter}) that use latent space perturbation. We perturb the latent variables \textbf{z$_{i}$} via Gaussian noise $\mathcal{N}(0,\alpha_{i})$ with a mean 0 and standard deviation \textbf{$\alpha$} along all dimensions \textbf{z} of the VED's latent space with width \textbf{$\mathrm{N}_{latent}$} (see equation 8). Let $\boldsymbol{\varepsilon}_{i}$ denote a sample vector from this $\mathrm{N}_{\text{latent}}$-dimensional Gaussian noise distribution. \textbf{$\alpha_{i}$} is a hyperparameter that controls the magnitude of the Gaussian noise added to each latent dimension. The resulting perturbed samples for each time step $t$ and grid cell $x$ are fed into the decoder of the VED to generate a stochastic parameterization (equation 8).

\begin{linenomath*}
\begin{equation}
\overline{\boldsymbol{Y}^{\mathrm{pred}}}=\frac{1}{N_{\mathrm{ens}}}\sum_{i=1}^{N_{\mathrm{ens}}}\boldsymbol{Y}_{i}^{\mathrm{pred}}\ \ ,\ \ \boldsymbol{Y}_{i}^{\mathrm{pred}}=\mathrm{VED}_{\boldsymbol{z}_{i}+\boldsymbol{\varepsilon}_{i}}\left(\boldsymbol{X}\right).
\label{eq:VED_stochastic}
\end{equation}
\end{linenomath*}

In equation~\ref{eq:VED_stochastic}, we create a stochastic parameterization by perturbing a single VED's latent space in two different ways: Either by adding isotropic Gaussian noise to the latent variables (``VED-static'', Tab. \ref{tab:overview_stoch_deter}) with $\alpha_i= 0.5$ to all latent variables \textbf{z$_{i}$}, or by adding anisotropic Gaussian noise whose standard deviation depends on the latent dimension (``VED-varying'', Tab. \ref{tab:overview_stoch_deter}). We evaluate the performance of these two stochastic parameterizations against a stochastic parameterization of the identical VED without latent space perturbation (``VED-draws'', Tab. \ref{tab:overview_stoch_deter}). Instead, VED-draws uses the repetitive draw from the latent space distribution based on the reparameterization (see section \ref{subsubsec:VED_theory}), where z is drawn from the latent space distribution based on $\mu$ and $\ln{\sigma^2}$. We show that VED-draws has limitations in the reproduction of convective processes and the representation of robust uncertainty quantification of them (Figs. {S10 - S16,S24}). In detail, the spread of VED-draws is considerably smaller compared to the two stochastic parameterizations with latent space perturbation. Therefore we do not show VED-draws in the following to simplify the visualization of our results.

We develop a thorough strategy for suitable $\alpha_{i}$ latent space perturbation. Its objective is to find a balance between reproduction skills and the ensemble spread of output predictions $\boldsymbol{Y}^{pred}$ by adjusting \textbf{$\alpha_{i}$} (see supporting information S.4 for details). In the following we show the applicability of the latent space perturbation approach tuned for one of the best-performing VEDs (VED 1, Tab. {S4}) and compare it against the other parameterization.

\ref{Appendix_A} briefly describes in section \ref{sec:Ens_metr} suitable ensemble metrics that we will apply to evaluate the skill of the developed stochastic and deterministic parameterizations. Additionally, \ref{Appendix_A} includes an explanation how we couple different DNNs to the numerical core of CESM2 in section \ref{sec:coupling}.

In the following section we evaluate the offline performance of our deterministic and stochastic parameterizations, while the online performance is presented in section \ref{sec:online}.

\section{Offline Evaluation: The Added Value of Multi-Member Parameterizations and Stochasticity \label{sec:Offline_Results}}

We evaluate the offline performance of our parameterization with three different types of metrics: 1) deterministic metrics to evaluate the reproduction of convective processes in section \ref{susbsec:deter_perf}, 2) metrics to investigate the quality of the uncertainty quantification in section \ref{subsec:Uncert_quant} and 3) the continuous rank probability score (CRPS) as a proper probabilistic score that allows an evaluation of reproduction and uncertainty calibration of our parameterizations in section \ref{subsec:CRPS}.  

\subsection{Deterministic Performance}
\label{susbsec:deter_perf}

\begin{figure}
    \centering
    \includegraphics[width=13cm]{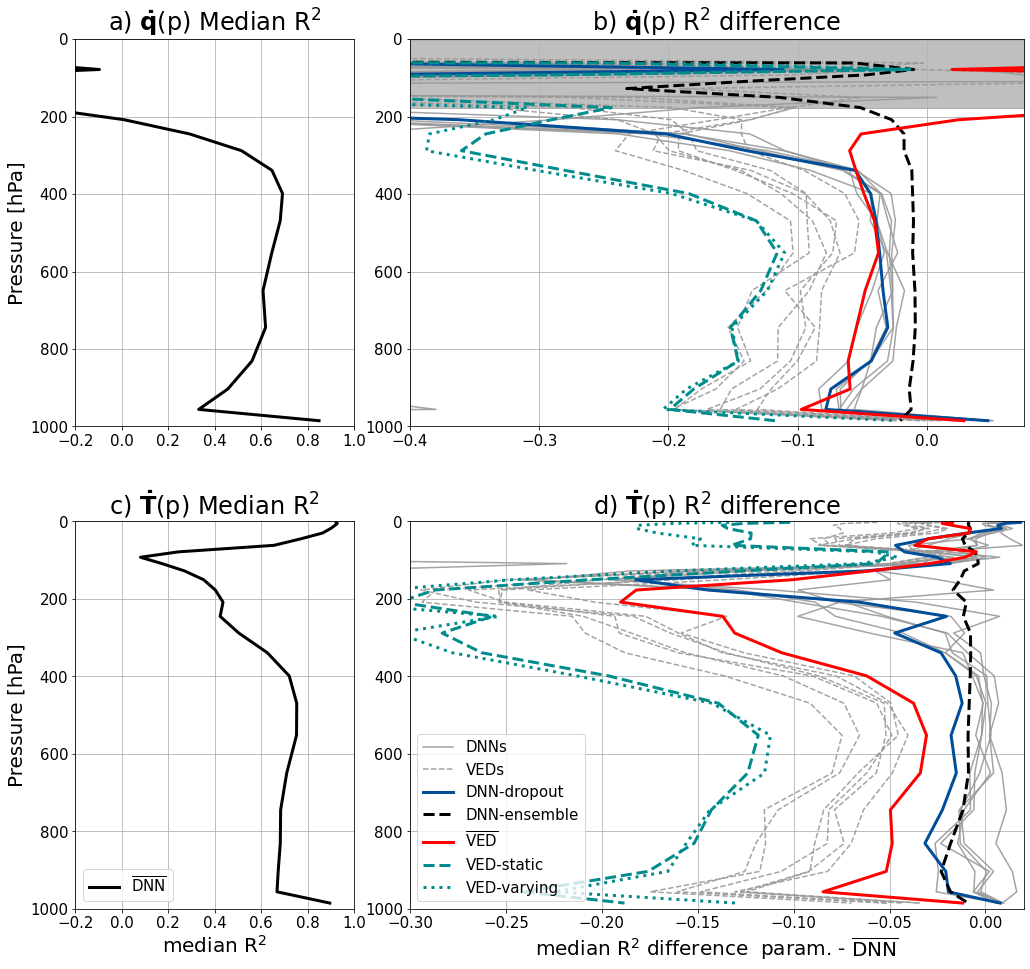}
    \caption{Vertical profiles of median coefficient of determination R$^2$ for specific humidity tendency $\boldsymbol{\dot{q}}$ of the mean predictions of the deterministic multi-member parameterization $ \mathrm{\overline{DNN}}$ (a), the differences of the median R$^2$ for $\boldsymbol{\dot{q}}$ of the mean predictions of DNN-dropout (solid navy blue); DNN-ensemble (dashed black), $\mathrm{\overline{VED}}$ (solid red); VED-static (dashed cyan) and VED-varying (dotted cyan line), as well as the predictions of individual DNNs and VEDs (grey solid and dashed lines) with respect to $ \mathrm{\overline{DNN}}$ (b), the median R$^2$ for temperature tendency $\boldsymbol{\dot{T}}$  of the mean predictions of $\mathrm{\overline{DNN}}$ in (c) and related differences of all other parameterizations in (d). The grey shaded area in (b) indicates the levels where the median R$^2$ of $\mathrm{\overline{DNN}}$ in (a) is below -0.05. The vertical profiles of median R$^{2}$ for cloud liquid water tendency $\boldsymbol{\dot{q}_{cl}}$ and cloud ice water tendency $\boldsymbol{\dot{q}_{ci}}$ can be found in Figure {S3}.}
    \label{fig:R_2_vert_profiles}
\end{figure}

We start our offline benchmark analysis by evaluating the reproduction performance of the different stochastic parameterizations compared to the deterministic multi-member parameterizations and individual DNNs and VEDs with respect to SP test data  (Tab. \ref{tab:overview_stoch_deter}). For this, we compute the coefficient of determination R$^2$ and the mean absolute error (MAE) along the time-dimension ($=$ 4020 time steps) in each of the grid cells and for all output variables $\boldsymbol{Y}$. For the stochastic and multi-member parameterizations we calculate these metrics based on the ensemble mean prediction for each time step and location. In the following we evaluate the median instead of the weighted mean of R$^2$ and MAE across all horizontal grid cells. One reason behind this is that both R$^2$ and MAE are not necessarily Gaussian. Moreover there is no variability in the test set in some grid cells for some output variables, e.g., $\boldsymbol{\dot{q}_{ci}}$ in the lower troposphere in the tropics, which limits the meaningfulness of grid averaged R$^2$ scores \cite{Yu2023_neurips}. 
Figure \ref{fig:R_2_vert_profiles} shows the median R$^2$ of the ensemble mean prediction of deterministic multi-member parameterization $\mathrm{\overline{DNN}}$ (Tab. \ref{tab:overview_stoch_deter}) across all horizontal grid cells for the vertical profiles of $\boldsymbol{\dot{q}}$ (Fig. \ref{fig:R_2_vert_profiles}a) and $\boldsymbol{\dot{T}}$ (\ref{fig:R_2_vert_profiles}c). The other panels indicate the differences of the median R$^2$ for the two profiles between all other parameterizations and $\mathrm{\overline{DNN}}$. This figure is complemented with a series of figures in the supporting information related to R$^2$ scores and MAEs with dedicated plots for all output variables $\boldsymbol{Y}$ and separation between land and ocean grid cells, to allow the reader a detailed analysis of the reproduction metrics of all DL models.

All DL models in Figure \ref{fig:R_2_vert_profiles} show an elevated reproduction skill for $\boldsymbol{\dot{T}}$ compared to $\boldsymbol{\dot{q}}$. The majority of models have a median R$^2$ $>$ 0.5 for these two tendency fields. All DL models have for $\boldsymbol{\dot{T}}$ and $\boldsymbol{\dot{q}}$ a higher median R$^2$ score over the ocean than over land in the free troposphere (Figs. S1,S2). 
Condensate tendencies $\boldsymbol{\dot{q}_{cl}}$ and $\boldsymbol{\dot{q}_{ci}}$ are more challenging to fit skillfully (Fig. {S3}), likely due to their small absolute magnitude as well as overall noisy and stochastic nature. For these vertical tendency profiles we see a median R$^2$ below 0.3 for all models. The median R$^2$ score for $\boldsymbol{\dot{q}_{ci}}$ is higher over land grid cells in the planetary boundary layer (Fig. S4), while over ocean grid cells the DL models have a higher median R$^2$ score for $\boldsymbol{\dot{q}_{cl}}$ near the phase transition layer in the mid troposphere on $\sim $ 500 hPa (Fig. S5). In Section \ref{sec:online} we will discuss the weaker offline performance for condensate tendencies associated with unstable CESM2 simulations, when condensate tendencies are included in the coupling. In general, DL models show a reproduction minimum in the lower troposphere and planetary boundary layer ($>$ 800 hPa, Figs. \ref{fig:R_2_vert_profiles}, S1 - S5,S7), due to the turbulent and stochastic nature of convective processes at these levels. The sharp decrease in performance in R$^2$ between the lower-most level (surface level) and the level above may reflect effects from surface coupling and the simulated turbulent processes with SP on coarse vertical grids \cite{Gentine2018,Pritchard2014b}.  The coefficient of determination indicates low reproduction skill above 200 hPa for the DL models for all variables except for $\boldsymbol{\dot{T}}$ (Fig. \ref{fig:R_2_vert_profiles}, {S1 - S5,S7}). However the related MAEs for $\boldsymbol{\dot{q}}$, $\boldsymbol{\dot{q}_{cl}}$, $\boldsymbol{\dot{q}_{ci}}$ above 200 hPa are almost null, as there is not much convection (Fig. {S11 - S13}). This underlines the fact that R$^2$ is not an optimal metric for the upper levels of the atmosphere with negligible convection at those levels \cite{Yu2023_neurips}. Despite this we acknowledge that stratospheric levels play a critical role with respect to stability of simulations with DL parameterizations when coupled to a host climate model, e.g., \citeA{Brenowitz2019,Kwa2023}. The related median MAE scores of the vertical profiles show the highest error on the surface levels for $\boldsymbol{\dot{q}}$ and $\boldsymbol{\dot{T}}$ for all DL models (Figs. S11 - S13). For $\boldsymbol{\dot{q}_{cl}}$ we see the highest median MAE within the upper planetary boundary layer, while for $\boldsymbol{\dot{q}_{ci}}$ it is located in the upper troposphere near 300 hPa.      

The advantages of the deterministic and stochastic multi-member parameterizations are immediately clear via the R$^2$ score analysis. In general, the ensemble mean of the deterministic DNN multi-member parameterization ($\mathrm{\overline{DNN}}$, Tab. \ref{tab:overview_stoch_deter}) and the ensemble mean of the stochastic DNN multi-member parameterization (DNN-ensemble, Tab. \ref{tab:overview_stoch_deter}) show an increased reproduction skill based on the R$^2$ scores for $\boldsymbol{\dot{q}}$ compared to single deterministic neural network predictions (grey lines in the background of Fig. \ref{fig:R_2_vert_profiles}b,d). However we find that the median MAE profiles of the ensemble mean of deterministic $\mathrm{\overline{DNN}}$ or stochastic DNN-ensemble parameterization have an in general weaker reproduction performance than individual DNNs due to a spurious member for the profiles of $\boldsymbol{\dot{q}}$ and $\boldsymbol{\dot{T}}$ (Figs. S11 - S13). Nevertheless clear advantages are present with lower median MAEs of DNN multi-member models over individual DNNs for the profiles of condensate tendencies (Figs. S11 - S13). $\mathrm{\overline{DNN}}$ and DNN-ensemble show nearly an equivalent performance for $\boldsymbol{\dot{T}}$, while the respective R$^2$ difference for $\boldsymbol{\dot{q}}$ ($\mathrm{\overline{DNN}}$ - DNNs) is larger than 0.02. 
In the lower troposphere one DNN has a slightly improved reproduction for $\boldsymbol{\dot{T}}$ compared to $\mathrm{\overline{DNN}}$ and DNN-ensemble.
The VED multi-member parameterization ($\mathrm{\overline{VED}}$, Tab. \ref{tab:overview_stoch_deter}) and the dropout-based DNN parameterization (DNN-dropout) result in enhanced reproduction skill compared to single VEDs, but these approaches are within the performance range of single DNNs. A similar skill of $\mathrm{\overline{VED}}$ and DNN-dropout compared to individual DNNs is visible for the median MAEs of $\boldsymbol{\dot{q}}$, $\boldsymbol{\dot{T}}$ and condensate tendencies (Figs. S11 - S13). A single VED with latent space perturbation (VED-static, VED-varying; Tab. \ref{tab:overview_stoch_deter}; Fig. \ref{fig:R_2_vert_profiles}b,d) show less reproductive capability than individual VEDs or VED-draws (without latent perturbation; not shown). The same decrease is also visible for the median MAEs of the vertical profiles (Figs. S11 - S13). We find that the median R$^2$ decays with increasing magnitude of the perturbation $\boldsymbol{\alpha_i}$ in initial experiments as expected (Fig. {S33}). This points to the fact that the magnitude of the latent space perturbation has to be well chosen to reach a good balance between reproduction skill and the diversity (ensemble spread) of the ensemble. We will see in the following that the perturbation of the latent space strongly improves the ensemble spread and can be well conditioned for a variety of output variables $\boldsymbol{Y}$.

The deterministic $\mathrm{\overline{DNN}}$ and stochastic DNN-ensemble multi-member parameterization improve the skill within the planetary boundary layer, which is a known challenge of DL subgrid parameterizations \cite{Gentine2018,Mooers2021,Behrens2022}. This is shown in Figure \ref{fig:R_2_vert_profiles}b, in which the minimal median R$^2$ for subgrid moistening $\boldsymbol{\dot{q}}$ in the boundary layer increases by more than 0.05 between individual DNNs and the deterministic multi-member parameterization $\mathrm{\overline{DNN}}$ or the stochastic multi-member parameterization DNN-ensemble. Figure S2 and Figure S3 indicate that the improved reproduction skill in the planetary boundary layer predominantly stems from the land area, where we have an in general higher reproduction (except Antarctica) compared to ocean grid cells.  Our analysis is based on the CESM native terrain following hybrid sigma-pressure grid. Therefore the shown pressure levels reflect always a reference pressure over the ocean based on the CESM reference pressure transformation.
To deepen the analysis, we focus now on the reference pressure level of $956~\mathrm{hPa}$ (the second level above the surface, Fig. S6). 
 The reproduction skill of $\boldsymbol{\dot{q}}$ in the planetary boundary layer is generally higher over land than over the ocean (Figs. S1,S2,S6) except for Antarctica for all evaluated DL models.  
The increase in reproduction skill of the deterministic $\mathrm{\overline{DNN}}$ and stochastic DNN-ensemble multi-member parameterization compared to individual DNNs is attributable to an improved representation of convective processes in the planetary boundary layer over Antarctica, the adjacent Southern Ocean and also over the Arctic Ocean (Fig. {S6}). The related improvement in R$^2$ with respect to DNN 1, as an example of a skillful DNN, or DNN-dropout exceeds in most grid cells in these regions more than 0.1 (Fig. S6).

There is no substantial added value of a stochastic or multi-member parameterization evident for precipitation rates and radiative fluxes for both median R$^2$ scores and median MAEs (Figs. S8 - S10 and S14 - S16). We see for most developed parameterizations reproduction capabilities (median R$^2 \ >\ 0.8$, see Fig. S{8}) comparable to reproduced 2D fields of single DNNs. Individual VEDs and the multi-member parameterization $\overline{\mathrm{VED}}$ tend to have the best reproduction skill for snow especially over ocean based on median R$^2$, while individual DNNs and DNN-dropout have the best performance for other 2D variables for both median R$^2$ and median MAE. Moreover $\overline{\mathrm{DNN}}$ and DNN-ensemble have a lower reproduction skill evaluated with median R$^2$ and median MAE for radiative and precipitation fluxes compared to almost all individual DNNs due to a member with almost no skill. However this member ensures online stability over a couple of months, as we will see in the following. For VED-varying and VED-static we see a similar and expected drop in reproduction skill for radiative and precipitation fluxes due to latent space perturbation. 
However, as we will see next, that the uncertainty quantification via latent space perturbation reflects a more calibrated magnitude with respect to SPCESM2 compared to other developed parameterizations.

\subsection{Evaluating Uncertainty Quantification}
\label{subsec:Uncert_quant}

Next, we evaluate the uncertainty quantification captured by the methods dealing with multiple predictions, meaning that prediction ranges from individual members will be assessed rather than their averages. We focus on four vertical subregions with larger than average MAEs (Figs. {S11 - S13}): $\boldsymbol{\dot{q}}$(p$_{surf})$; $\boldsymbol{\dot{T}}$(p$_{surf})$; $\boldsymbol{\dot{q}_{cl}}$(800-900 hPa); and $\boldsymbol{\dot{q}_{ci}}$(200-400 hPa). These levels correspond also to the levels with strong variability in the test data. Therefore these levels reflect the most pronounced biases in uncertainty quantification with respect to the test data of our parameterizations. We choose this focus on levels with large biases in uncertainty quantification to critically evaluate the fit-for-purpose of our parameterizations. The following figures serve as a test-bed for how well the developed multi-member and stochastic parameterizations capture uncertainties in such a challenging environment. 

Figure \ref{fig:spread_skill} shows the spread-skill diagrams (section \ref{sec:Ens_metr} in \ref{Appendix_A}) for surface $\boldsymbol{\dot{q}}$ and $\boldsymbol{\dot{T}}$. An ideal spread-skill ratio of 1 is indicated by the grey dashed line \cite{Berner2017}. We randomly draw 500 time steps from the test set ($\sim 6.9\times 10^6 $ samples), and calculate the spread. Then we bin the spread arrays into 41 bins, based on the spread percentiles of VED-static, with bin widths of 2.5$^{th}$ percentiles. We finally calculate the conditional average of spread and  Root Mean Squared Error (RMSE) for each bin (equation \ref{eq:RMSE_def}). The y-axis and x-axis represent the bin-averaged RMSE and spread, respectively. To put the magnitude of the shown maximum spread and RMSE values into perspective, their values are typically $10^2$ to $10^3$ larger than the MAEs (Figs. {S11 - S13}). The spread-skill analysis is complemented with an analysis of the corresponding probability integral transform (PIT, section \ref{sec:Ens_metr} in \ref{Appendix_A}) histograms. Figure \ref{fig:PIT_dq_cld_liq_dt} shows the PIT histogram for $\boldsymbol{\dot{q}_{cl}}$ in the planetary boundary layer and further related PIT histograms can be found in the supporting information. The ideal PIT curve is shown as the thick dashed grey line in Figure \ref{fig:PIT_dq_cld_liq_dt}.

\begin{figure}[h!t]
    \centering
    \includegraphics[width=13cm]{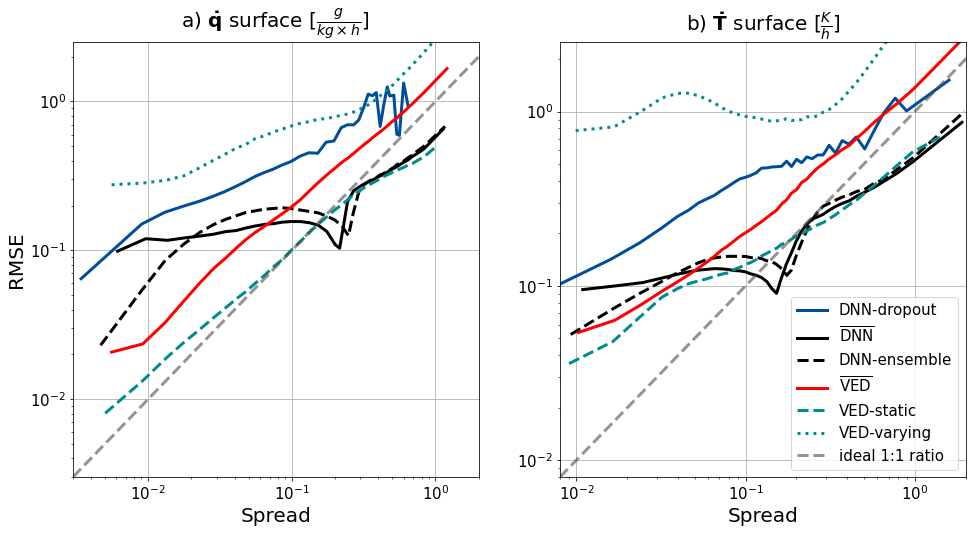}
    \caption{Spread-Skill diagram between bin-averaged spread (x-axis) and Root Mean Square Error (RMSE, y-axis) based on the test data and predictions over 500 randomly drawn timesteps. Shown is the spread-skill diagram of surface specific humidity tendency $\boldsymbol{\dot{q}}$ in a) and surface temperature tendency $\boldsymbol{\dot{T}}$ in b). The color-coding of the multi-member and stochastic parameterizations is identical to Fig. 2. Additionally we include the spread-skill ratio of 1:1 (dashed grey line) that symbolises the optimal calibration of the spread vs. skill based on literature \cite{Berner2017,Haynes2023}.}
    \label{fig:spread_skill}
\end{figure}

We find the best performance with respect to the spread-skill diagrams for VED-static followed by the stochastic DNN-ensemble and  deterministic $\overline{\mathrm{DNN}}$ multi-member parameterization. As it is shown in Figure \ref{fig:spread_skill}, for a spread smaller than 0.35 $\frac{g}{kg \times h}$ or $\frac{K}{h}$ for surface $\boldsymbol{\dot{q}}$ or $\boldsymbol{\dot{T}}$, these three parameterizations provide a considerably skillful uncertainty quantification. For larger spreads of surface $\boldsymbol{\dot{q}}$ and $\boldsymbol{\dot{T}}$ VED-static, DNN-ensemble and $\overline{\mathrm{DNN}}$ illustrate an overdispersion (underconfidence, where the ensemble overestimates the variations in $Y_{j}$). This means that the associated spread is larger than the RMSE and the respective spread-skill curves are situated below the ideal 1:1 ratio line. While for $\boldsymbol{\dot{q}_{cl}}$ in the planetary boundary layer and $\boldsymbol{\dot{q}_{ci}}$ in the upper troposphere, the underdispersion (overconfidence, where the ensemble underestimates the variations in $Y_{j}$), when the spread is smaller than the RMSE, reduces with $\mathrm{\overline{DNN}}$, DNN-ensemble and VED-static compared to all other developed parameterizations (Fig. S17). To deepen the analysis we compute also spread-skill diagrams for ocean and land grid cells (Figs. S18,S19), but did not find considerable land-sea contrasts in spread-skill diagrams. This suggests an improved uncertainty calibration for VED-static and $\overline{\mathrm{DNN}}$, which is also found in the respective PIT curves (Fig. \ref{fig:PIT_dq_cld_liq_dt}). Figure \ref{fig:PIT_dq_cld_liq_dt} shows the PIT diagram for $\boldsymbol{\dot{q}_{cl}}$ in the planetary boundary layer (Fig. \ref{fig:PIT_dq_cld_liq_dt}). The ideal PIT curve is shown as the thick dashed grey line. VED-static and  $\overline{\mathrm{DNN}}$ are in general closest to the ideal PIT curve. They show either only weak overdispersion or underdispersion for all evaluated variables for both land and ocean grid cells (Figs. S20 - S23). It has to be noted that the differences between land and ocean grid cells are subtle like for the spread-skill diagrams. For the stochastic DNN-ensemble multi-member parameterization we see a decreased quality of the uncertainty quantification based on the PIT curves with too heavy tails and a too dominant central rank for all evaluated variables.

DNN-dropout yields less calibrated uncertainty quantification, with larger deviations from the ideal 1:1 ratio and strong underdispersion for the evaluated variables in spread-skill diagrams compared to the other approaches (Figs. 3, S17 - S19). The pronounced underdispersion is also present in the associated PIT diagrams (Figs. \ref{fig:PIT_dq_cld_liq_dt}, S20 - S23). For DNN-dropout, almost all test data samples are situated in the tails of the distribution of the PIT curve for all variables for both ocean and land grid cells. 
In combination with the overall poor skill in the spread-skill diagrams (Fig. \ref{fig:spread_skill}), except for $\boldsymbol{\dot{T}}$ for DNN-dropout, this suggests that the DNN-dropout yields an uncertainty quantification that underestimates the variability in the test data for all evaluated variables.
In the following, we will show how this translates into a poor CRPS skill for DNN-dropout.  

Similarly, $\overline{\mathrm{VED}}$ tends to be underdispersive for all evaluated variables (Figs. \ref{fig:spread_skill},\ref{fig:PIT_dq_cld_liq_dt}, S17 - S23), but with an improved spread-skill relationship and weaker underdispersion in PIT curves compared to DNN-dropout. Also we find that $\overline{\mathrm{VED}}$ is competitive against all other parameterizations for smaller spread values (Fig. \ref{fig:spread_skill}). 
This suggests that the $\mathrm{\overline{VED}}$ provides better calibrated uncertainty quantification compared to DNN-dropout. In the following the CRPS evaluation will further support this reasoning.

\begin{figure}[ht]
    \centering
    \includegraphics[width=13cm]{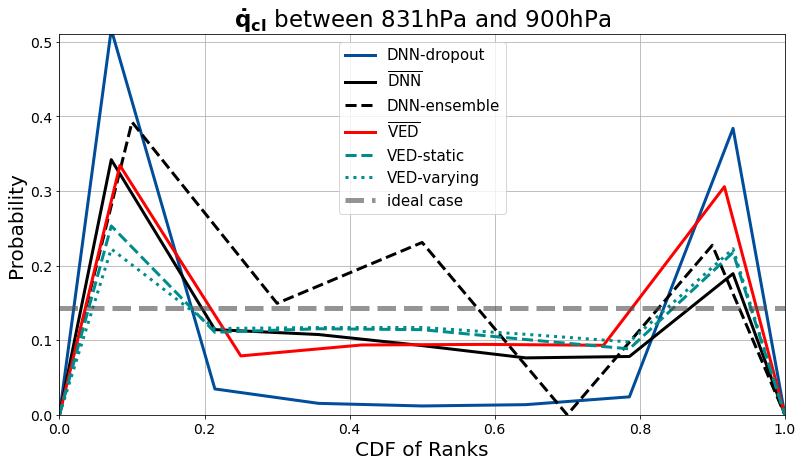}
    \caption{Probability Integral Transform (PIT) histogram of $\boldsymbol{\dot{q}_{cl}}$ in the planetary boundary layer between 831 and 900 hPa. The x-axis represents the cumulative distribution function (CDF) of the ranks of the test sample from SPCESM2 with respect to the number of ensemble members of the stochastic or multi-member parameterizations. The y-axis depicts the probability associated with each rank. The PIT histogram is based on 400 randomly drawn time steps from the test set. The thick dashed grey line in horizontal direction symbolises the ideal shape of the PIT curve.}
    \label{fig:PIT_dq_cld_liq_dt}
\end{figure}

The spread-skill analysis reveals substantial differences in the quality of the uncertainty quantification between a latent space perturbation with isotropic Gaussian noise (VED-static) and anisotropic Gaussian noise (VED-varying). While VED-static is one of the best performing parameterizations, we find a pronounced underdispersion for VED-varying for $\boldsymbol{\dot{q}}$ and $\boldsymbol{\dot{T}}$ at the surface (Fig. \ref{fig:spread_skill}), more so than for the DNN-dropout. This result of the spread-skill analysis is particularly interesting as it suggests that a latent space perturbation with an anisotropic Gaussian noise term (VED-varying) yields a decreased calibration of the uncertainty quantification of the surface moistening and heating compared to an isotropic Gaussian noise term (VED-static). However for $\boldsymbol{\dot{q}_{cl}}$ in the planetary boundary layer and $\boldsymbol{\dot{q}_{ci}}$ in the upper troposphere both VED-varying and VED-static show an improved calibration of the ensemble spread compared to all other developed multi-member and stochastic parameterizations for a bin averaged spread larger than 0.005 $\frac{g}{kg \times h}$ (Fig. S17). VED-varying shows a weaker prediction skill compared to VED-static for $\boldsymbol{\dot{q}_{cl}}$ and $\boldsymbol{\dot{q}_{ci}}$ for a spread smaller than 3 $\times 10^{-4}~\frac{g}{kg \times h}$ (Fig. \ref{fig:spread_skill}). This results in an increased underdispersion of VED-varying compared to DNN-dropout for a spread smaller than 2 $\times 10^{-4}~\frac{g}{kg \times h}$. We could then cross-link the results from the spread-skill diagrams of VED-static and VED-varying with the respective PIT histograms (Figs. \ref{fig:PIT_dq_cld_liq_dt}, S20 - S23). Figure \ref{fig:PIT_dq_cld_liq_dt} shows that VED-static and VED-varying have strongly reduced outliers in their respective PIT histograms for $\boldsymbol{\dot{q}_{cl}}$ in the planetary boundary layer. The calibration of the uncertainties for VED-varying is slightly improved compared to that of VED-static. The probability that the true SPCESM2 sample is ranked at the outer edge of the PIT curves decreases for VED-varying, while the probabilities for the inner ranks for VED-varying converge towards the ideal case for $\boldsymbol{\dot{q}_{cl}}$ in the planetary boundary layer (Fig. \ref{fig:PIT_dq_cld_liq_dt}). The same improved quality of uncertainty quantification is also present for $\boldsymbol{\dot{q}}$, $\boldsymbol{\dot{T}}$ at the surface and $\boldsymbol{\dot{q}_{ci}}$ in the upper troposphere with VED-varying compared to VED-static (Figs. {S20 - S23}). However, we recall that this comes at the cost of worse predictive skill of convective processes based on ensemble mean predictions (Figs. \ref{fig:R_2_vert_profiles},\ref{fig:spread_skill}).

Overall we find that the stochastic VED-static parameterization has the best uncertainty quantification on levels with pronounced variability and thus general underdispersion based on the PIT curves and the spread-skill diagrams. It is followed by the determinisitic $\mathrm{\overline{DNN}}$ multi-member parameterization with a good calibration of the ensemble spread. These networks often indicate only a slight underdispersion or overdispersion compared to the ideal PIT curve. The stochastic DNN-ensemble multi-member parameterization has a good performance with respect to the spread-skill ratio, but has larger deviations from the ideal PIT curve than VED-static or $\mathrm{\overline{DNN}}$. The stochastic VED-varying parameterization provides calibrated uncertainty quantification in the PIT analysis but to the expense of a low reproduction skill as can be seen in its large RMSE for $\boldsymbol{\dot{q}}$, $\boldsymbol{\dot{T}}$ at the surface and condensate tendencies (Figs. \ref{fig:spread_skill}, S17 - S19). The $\mathrm{\overline{VED}}$ multi-member parameterization represents the uncertainty of convective processes with intermediate results in spread-skill diagrams and PIT curves. The uncertainty quantification of the stochastic DNN-dropout parameterization is not calibrated and exhibits the strongest underdispersion of all developed parameterizations. Most of the true SPCESM2 samples are sorted in the lowest or highest rank as outliers in the PIT curves of DNN-dropout. This means that DNN-dropout strongly underestimates the simulated spread of key variables in SPCESM2, and can not represent variations in convective processes like all other parameterizations.

\subsection{Proper Scoring}
\label{subsec:CRPS}

\begin{figure}
    \centering
    \includegraphics[width=\textwidth]{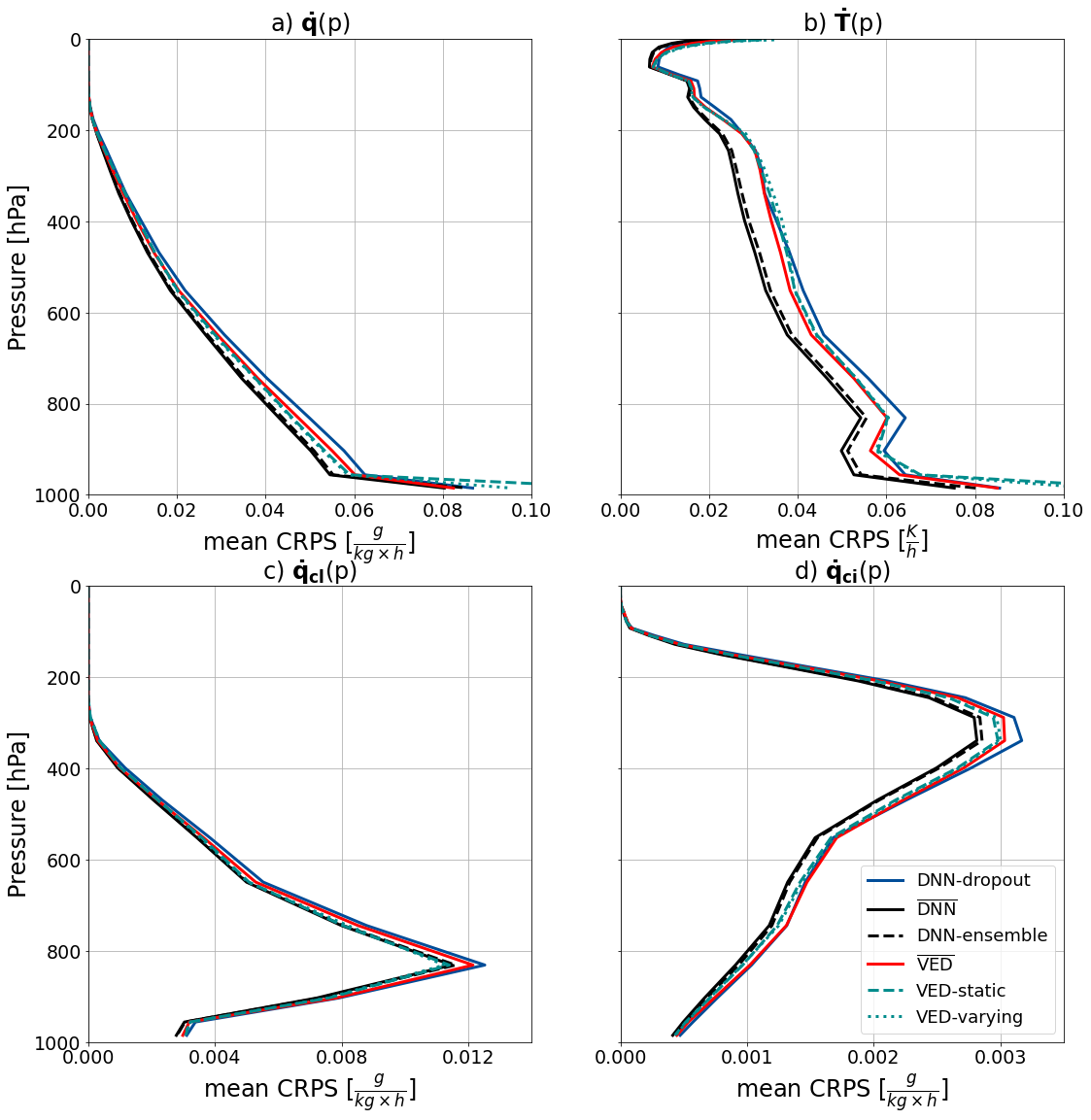}
    \caption{Mean Continuous Rank Probability Score (CRPS) of $\boldsymbol{\dot{q}}$ (a), $\boldsymbol{\dot{T}}$ (b), $\boldsymbol{\dot{q}_{cl}}$ (c), $\boldsymbol{\dot{q}_{ci}}$ (d) for the different ensembles based on 500 randomly drawn time steps from the test data. The blue line indicates DNN-dropout. The solid and dashed black lines represent the deterministic $\overline{\mathrm{DNN}}$ and stochastic DNN-ensemble parameterization alongside $\overline{\mathrm{VED}}$ (red line). The dashed and dotted cyan lines represent VED-static and VED-varying.}
    \label{fig:mean_crps_values_ensemble}
\end{figure}

Here we provide a holistic evaluation of both the calibration of the ensemble spread and the quality of the reproduction based on the continuous rank probability score (CRPS, see section \ref{sec:Ens_metr} in \ref{Appendix_A}). $\overline{\mathrm{DNN}}$ and DNN-ensemble are the best-performing deterministic and stochastic parameterization based on CRPS (Fig. \ref{fig:mean_crps_values_ensemble}). We start our CRPS analysis by focusing on the general statistics of CRPS calculated over all output variables $\boldsymbol{Y}$. 
We find the lowest mean and median CRPS of $\boldsymbol{Y}$ for the deterministic $\overline{\mathrm{DNN}}$ and stochastic DNN-ensemble multi-member parameterization (Fig. S24). This indicates that these two parameterizations are the best compromise between predictive skill on one side and uncertainty quantification on the other side. While $\overline{\mathrm{VED}}$ and DNN-dropout perform considerably well, VED-draws shows intermediate performance based on the mean and higher percentiles of CRPS calculated over $\boldsymbol{Y}$. Both VED-static and VED-varying have remarkably increased 75$^{th}$ and 90$^{th}$ percentiles compared to all other parameterizations (Fig. {S24}). However, we note that the respective median CRPS decrease compared to the initial stochastic VED-draws parameterization (section \ref{subsec:lat_perturb}, Tab. \ref{tab:overview_stoch_deter}), which underscores that the latent space perturbation has the potential to improve the uncertainty quantification of convective processes. As a next step we computed the mean over all grid cells, the ocean grid cells, and land grid cells exclusively of vertically averaged CRPS for $\boldsymbol{\dot{q}\left(p\right)}$, $\boldsymbol{\dot{T}\left(p\right)}$, $\boldsymbol{\dot{q}_{cl}\left(p\right)}$ and $\boldsymbol{\dot{q}_{ci}\left(p\right)}$ (Figs. S25 - S28). For all parameterizations we see higher CRPS over ocean grid cells than over land. This indicates that the predictive skill of our parameterizations tends to be higher over land than over ocean grid cells, or from a data science perspective that it is more challenging to learn convective processes over the ocean than over the land. Interestingly, the differences in CRPS between the different parameterizations show a similar behaviour for both ocean and land points (Figs. S25 - S28). For all profiles $\overline{\mathrm{DNN}}$ and DNN-ensemble have the lowest CRPS. The two stochastic parameterizations based on latent perturbation show considerable improvements for cloud condensate tendencies, while elevated CRPS for $\boldsymbol{\dot{q}}$ and $\boldsymbol{\dot{T}}$ compared to all other parameterizations. DNN-dropout or $\overline{\mathrm{VED}}$ have high or average CRPS but without the variations in CRPS between variables that we see for VED-static and VED-varying (Figs. S25 - S28).

We extend our CRPS analysis to evaluate from which model levels the differences between the parameterizations are arising. Figure \ref{fig:mean_crps_values_ensemble} shows the vertical profiles of mean CRPS values for $\boldsymbol{\dot{q}\left(p\right)}$, $\boldsymbol{\dot{T}\left(p\right)}$, $\boldsymbol{\dot{q}_{cl}\left(p\right)}$ and $\boldsymbol{\dot{q}_{ci}\left(p\right)}$. 
The similar performance in mean CRPS of $\overline{\mathrm{DNN}}$ and DNN-ensemble suggests that the latter as a stochastic parameterization does not exhibit a decline in reproduction skill of convective processes, as found with all other developed stochastic parameterizations, particularly in the upper planetary boundary layer and the upper troposphere.
VED-static and VED-varying have a compatible performance to $\overline{\mathrm{DNN}}$ and DNN-ensemble in the upper part of the planetary boundary layer for $\boldsymbol{\dot{q}_{cl}}$ and in general a good skill for all vertical profiles (Fig. \ref{fig:mean_crps_values_ensemble}). However VED-static and VED-varying have an elevated CRPS for surface $\boldsymbol{\dot{q}}$ and $\boldsymbol{\dot{T}}$ compared to other deterministic or stochastic parameterizations. 
The shortcomings of VED-static and VED-varying on the surface levels mainly arise from the reduced reproduction skill with latent space perturbation while the calibration of the uncertainty quantification depicts high skill (Figs. {S22,S23}). $\overline{\mathrm{VED}}$ shows in general a compatible performance in CRPS with intermediate scores, while DNN-dropout depicts the highest CRPS of all evaluated parameterizations for all vertical profiles due to the shortcomings in the calibration of the ensemble spreads (Figs. \ref{fig:spread_skill},\ref{fig:mean_crps_values_ensemble}).

\begin{figure}[t]
    \centering
    \includegraphics[width=\textwidth]{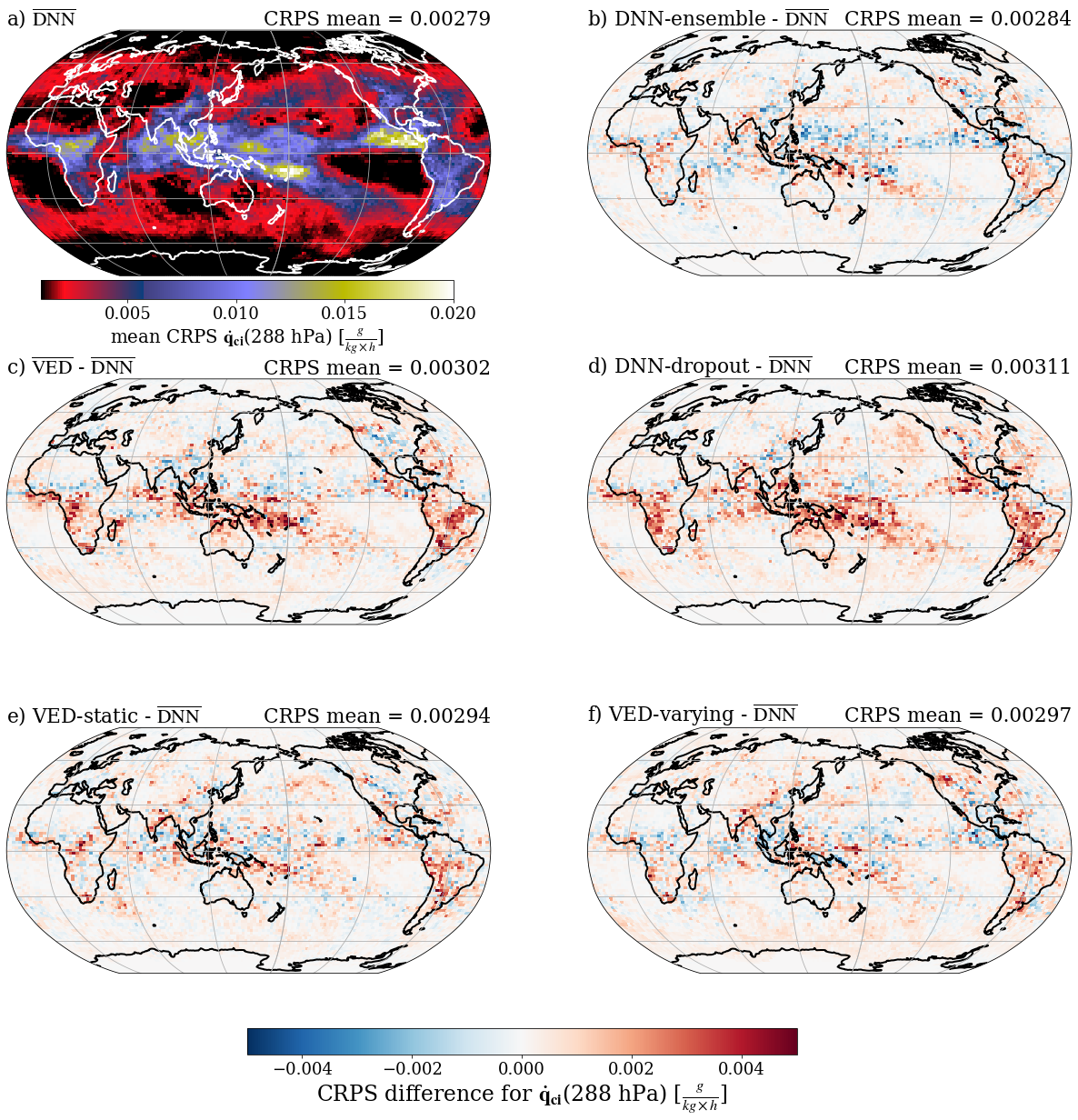}
    \caption{Mean Continuous Rank Probability Score (CRPS) of $\boldsymbol{\dot{q}_{ci}}$ on 288 hPa based on 500 randomly drawn timesteps from the test set for (a) the deterministic $\overline{\mathrm{DNN}}$, the CRPS differences of (b) the stochastic DNN-ensemble, (c) the deterministic $\overline{\mathrm{VED}}$, (d) DNN-dropout; VED-draws (e), VED-static (f), VED-varying (g) parameterizations to $\overline{\mathrm{DNN}}$. The CRPS mean over the global map for each parameterization is printed above each panel in the right top corner.}
    \label{fig:CRPS_CLDICEBCTEND_288}
\end{figure}

 Figure \ref{fig:CRPS_CLDICEBCTEND_288} shows the global map of the mean CRPS values of $\boldsymbol{\dot{q}_{ci}}$ on 288 hPa for $\mathrm{\overline{DNN}}$ based on 500 randomly drawn time steps from the test set. Moreover it depicts the differences of mean CRPS of all other developed parameterizations with respect to $\mathrm{\overline{DNN}}$. 
 In the supporting information similar maps for $\boldsymbol{\dot{q}_{cl}}$, surface $\boldsymbol{\dot{q}}$ and $\boldsymbol{\dot{T}}$ (Fig. {S29 - S31}) can be found. The CRPS structure shows the imprint of the atmospheric general circulation centers of action. In general, we find the largest mean CRPS, a lower reproduction skill of our parameterizations, in the ITCZ region like especially over the Maritime continent or over the tropical East Pacific offshore of Panama (Figs. \ref{fig:CRPS_CLDICEBCTEND_288},S29,S30). For the CRPS of surface $\boldsymbol{\dot{T}}$ the lowest skill is mostly visible over the tropical arid and semiarid regions, like Southern Africa, the Sahel region and Western Australia (Fig. S31). 
Especially over these regions with high CRPS for all evaluated levels and variables $\mathrm{\overline{DNN}}$ and DNN-ensemble have the best performance compared to the other parameterizations (Figs. \ref{fig:CRPS_CLDICEBCTEND_288}, S29 - S31). For $\boldsymbol{\dot{q}_{ci}}$ on 288 hPa $\mathrm{\overline{DNN}}$ and DNN-ensemble have the lowest global mean CRPS with 2.8$\times 10^{-3}~\frac{g}{kg \times h}$, while the other parameterizations have a mean value larger than 3$\times 10^{-3}~\frac{g}{kg \times h}$ except for VED-static and VED-varying (Fig. \ref{fig:CRPS_CLDICEBCTEND_288}). DNN-dropout has in general elevated CRPS over the deep convective regions for $\boldsymbol{\dot{q}_{ci}}$ compared to the other developed parameterizations due to its strong underdispersion (Fig. S21). Therefore it is not surprising that VED-static and VED-varying have a lower mean CRPS for upper tropospheric $\boldsymbol{\dot{q}_{ci}}$ due to their improved quality of uncertainty estimates compared to DNN-dropout.

The extra-tropical and especially regions with negligible deep convective activity, i.e. the upwelling regions offshore of the west coast of the Americas or Africa, are characterized by similar small CRPS across all parameterizations, as expected (Figs. \ref{fig:CRPS_CLDICEBCTEND_288}, S29 - S31). In agreement with previous results, DNN-dropout often has elevated CRPS. For VED-static and VED-varying we find an improvement in CPRS compared to $\overline{\mathrm{DNN}}$ and DNN-ensemble for $\boldsymbol{\dot{q}_{cl}}$ on 831 hPa, but the largest CRPS for surface $\boldsymbol{\dot{q}}$ and surface $\boldsymbol{\dot{T}}$ as already expected from Figure \ref{fig:mean_crps_values_ensemble}.

 In general, the latent space perturbation leads to an improvement in the calibration of the ensemble spread compared, for example, to DNN-dropout. Nevertheless, our CRPS and the uncertainty analysis reveal that there is a trade-off between robust uncertainty quantification on one hand and reproduction skill on the other hand. Therefore we designed a hyperparameter tuning method to balance these two important factors for the development of a stochastic convection parameterization with latent space perturbation (see SI section {S.4}).
 The individual members of DNN-ensemble and $\overline{\mathrm{DNN}}$, do not need such additional tuning steps and show a similarly good calibration of the uncertainty quantification of convective processes in combination with enhanced reproduction skill of convective processes compared to all other developed parameterizations.

In the next section, we will therefore focus on $\overline{\mathrm{DNN}}$ and DNN-ensemble parameterizations coupled to CESM2, demonstrating the advantages of such parameterizations compared to single DNN parameterizations.

\section{Online Results: Improved Stability and Precipitation Distributions\label{sec:online}}

In this section, we first describe the challenges of coupling our parameterizations to CESM2. Second, we evaluate our prognostic runs against the high-resolution SPCESM2 model, the coarse CESM2 model with a traditional convection scheme and precipitation observations, as well as other deep learning benchmarks. We acknowledge that our online simulations have to be interpreted as experiments and not as an operational setup due to our reliance on the superparameterization for condensate tendency emulation. This has the clear disadvantage of slower CESM2 simulations than with the original superparameterization (see Tab. S6). Moreover we note that the advantages in the reproduction of tropical extreme precipitation illustrated in subsection \ref{sub_sec:online_runs}  cannot be fully attributed to our multi-member parameterizations due to the reliance on an SP call for condensate tendencies.

\subsection{Online Coupling Challenges}

We couple $\overline{\mathrm{DNN}}$ and DNN-ensemble, the two best-performing deterministic and stochastic parameterizations, into CESM2 using the Fortran-Keras-Bridge \cite{Ott2020}, resulting in $\overline{\mathrm{DNN}}$-SP-CESM and DNN-ens-SP-CESM hybrid models. ``Hybrid model'' means in this context that we couple a DL algorithm to the numerical core of a climate model or ESM.  We follow the configuration detailed in section \ref{sec:coupling} in \ref{Appendix_A} for our new hybrid model runs. Coupling the complete set of $\boldsymbol{Y^{pred}}$ to CESM2 led to unstable prognostic runs after few days. Note that running the hybrid model with an individual DNN led to instabilities in only a few time steps. This shows the stabilizing effect of ensembles consistent with \citeA{Brenowitz2020}. We identified one particular DNN with low performance of the parameterizations and retrained it. This allowed us to achieve longer prognostic runs of a few weeks. While the stability of the prognostic runs depends on the initial conditions, the primary cause of the hybrid model instabilities were ice growth in the lower tropical stratosphere and subsequent radiative feedback. These anomalous signals manifested in rapidly increasing $\boldsymbol{q_{ci}}$ in the stratosphere, which led to unrealistic values of $\boldsymbol{Y^{pred}_{rad}}$ that are crucial for the coupling with the surface model components (e.g. land model), ultimately causing blow-ups of CESM2. 

Achieving a stable hybrid multi-scale model is a non trivial task \cite{Yu2023_neurips}. The deep learned representation of condensate tendencies is particularly challenging in CESM2. \citeA{Yuval2021} overcame this issue by constructing one neural network emulating solely surface subgrid fluxes and one neural network dedicated to predicting the tendencies terms in the free atmosphere including condensate tendencies in the System for Atmospheric Modeling in aquaplanet simulations. Recently \citeA{Yu2024} and \citeA{Hu2024} showed a novel DL framework with stable online coupling of cloud condensate tendencies for a different ESM. However to overcome this challenge of cloud condensate tendencies in CESM2, we performed a ``perfect condensate'' experiment, in which $\boldsymbol{\dot{q}_{ci}}$ and $\boldsymbol{\dot{q}_{cl}}$ variables are simulated by the SP component and the rest by our deep learned parameterizations. This partially-coupled setup, however, requires running the SP component alongside the predictions from the neural networks, with a clear drawback in terms of computational efficiency (Tab. S6). Nevertheless, this configuration allowed us to achieve five months of stable hybrid runs for both $\overline{\mathrm{DNN}}$-SP-CESM and DNN-ens-SP-CESM. Specifically, $\overline{\mathrm{DNN}}$ coupled to CESM ran from the beginning of February 2013 to the last third of July, and DNN-ensemble coupled to CESM stopped at the beginning of July. Both runs failed due to a strong temperature decrease at the surface in the tropics, which was driven by a pronounced surface radiative imbalance. The resulting temperature decrease finally caused a violation of the predefined temperature thresholds of CESM2 (Figs. {S35,S36}). 

Running the hybrid model with the ``perfect condensate'' setup but for individual DNNs, crashed in six out of seven cases within the first five days of the simulation ({see Figs. S37,S38}). The DNN with the largest RMSE due to imperfect predictions representing average conditions (e.g. predicting constant drizzle conditions in all horizontal grid cells) survived until mid October ({Figs. S37,S38}). The respective DNN showed already poor performance in our offline reproduction analysis (section \ref{sec:Offline_Results}), but had one of the best training skills in the initial hyperparameter tuning stage. This suggests that model stability and the robustness or realism of the predicted convective and radiative fluxes are not associated with each other \cite{Lin2023}. Omitting the spurious DNN as a member destabilized the hybrid simulations with $\overline{\mathrm{DNN}}$ and DNN-ensemble in test runs. This sanity check shows that there exists also a vital trade-off between online stability and induced biases due to imperfect DL predictions that we will see in the following. Furthermore, we found that using fewer members (number of neural networks and larger stochasticity) for DNN-ens-SP-CESM strongly affected its stability. As an example, DNN-ens-SP-CESM with two members crashed within the first month of simulation. This shows that deep-learned multi-member parameterizations require a trade-off between computational efficiency and the number of members (see the differences in computational requirements between $\overline{\mathrm{DNN}}$-SP-CESM and DNN-ens-SP-CESM in Tab. 6). We further performed experiments with different initialisation dates (January, June and October 2013) and found that all developed multi-member parameterizations are unstable within the first month. This suggests that model stability may well depend also on the seasonality governing ESM simulations. 

\subsection{Online Performance}
\label{sub_sec:online_runs}

\begin{figure}
    \centering
    \includegraphics[width=12cm]{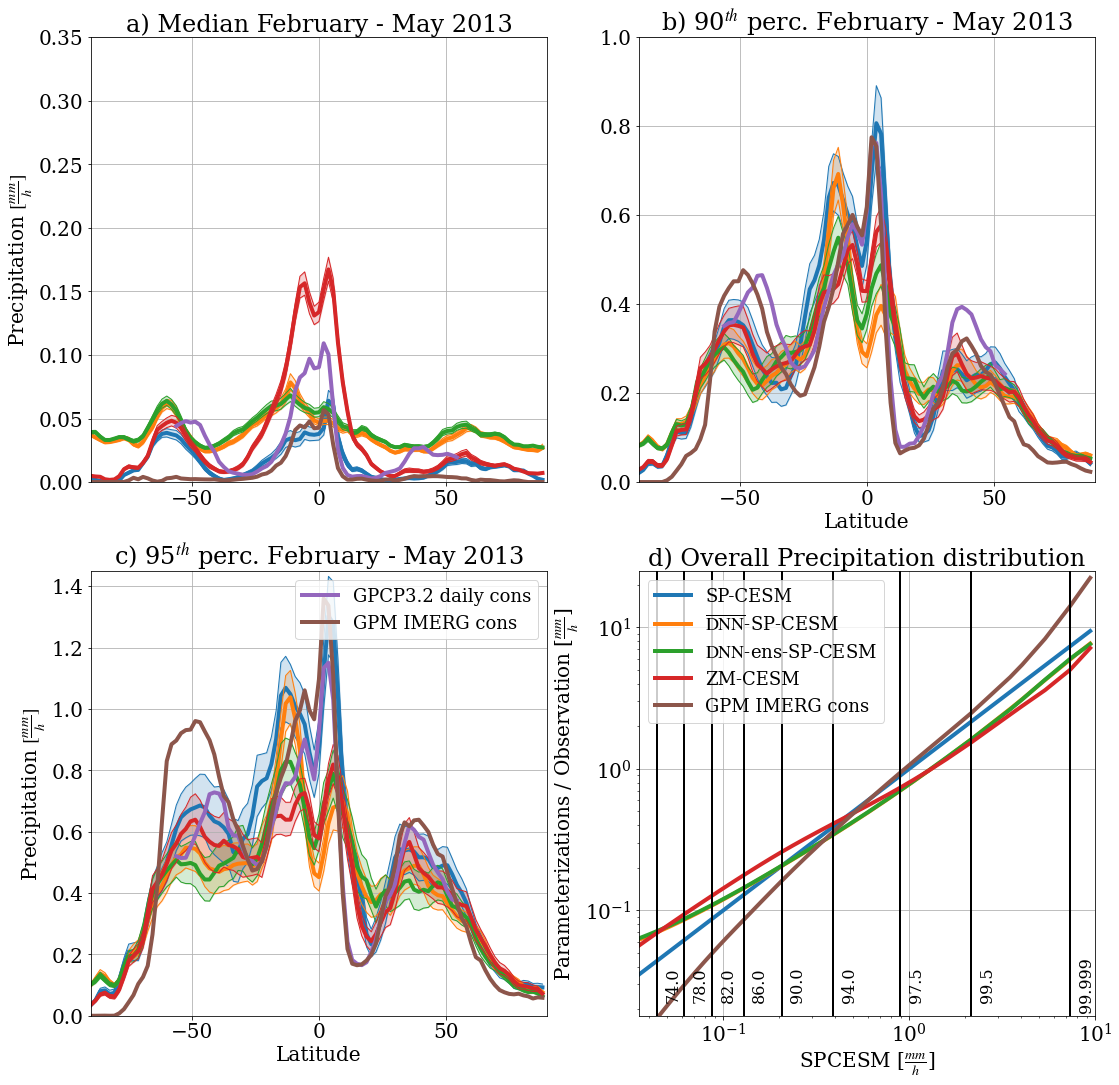}
    \caption{Simulated zonal averages of median (a), 90$^{th}$ (b) and 95$^{th}$ percentiles (c) of total precipitation in the period February to May 2013 of CESM2 with a superparameterization (SP-CESM, blue), CESM2 with the deterministic $\overline{\mathrm{DNN}}$ parameterization ($\overline{\mathrm{DNN}}$-SP-CESM, orange), CESM2 with the stochastic DNN-ensemble parameterization (DNN-ens-SP-CESM, green) and CESM2 with the traditional Zhang-McFarlane scheme (ZM-CESM, red line). The uncertainty ranges for the CESM2 simulations indicate the span between minimum and maximum of the given metrics based on bootstrapping with 50 subsamples due to internal variability. Additionally the zonal averages of median, 90$^{th}$ and 95$^{th}$ percentiles are displayed of the two observational datasets GPCP3.2 with daily resolution (GPCP3.2 daily cons, purple) and GPM IMERG (GPM IMERG cons, brown lines). Subplot d) shows the precipitation distribution  of the different parameterizations or observation (y-axis) as a function of the precipitation distribution simulated with the superparameterization (x-axis). The vertical lines in subplot d) represent distinct percentiles of the precipitation distribution in SP-CESM. For subplot d) the entire simulated precipitation rates in all grid cells and all time steps of the period February to May 2013 are used.}
    \label{fig:online_precip_distrib}
\end{figure}

We evaluate $\overline{\mathrm{DNN}}$-SP-CESM and DNN-ens-SP-CESM prognostic runs over the period from February to May 2013 for the precipitation analysis and from February to July 2013 for the large-scale temperature and specific humidity fields (before blow-up in mid-July). We chose the shorter period for the evaluation of precipitation due to its strong seasonality that influences diurnal cycles for example. 
These simulations are evaluated against the original high-resolution SPCESM2 (abbreviated as SP-CESM), and against the coarse CESM2 (abbreviated as ZM-CESM) with the traditional convection parameterization \cite{ZhangMcFarlane1995}, over the same period. Moreover we use daily data from the Global Precipitation Climatology Project version 3.2 (GPCP3.2 daily cons, Tab. S7, \citeA{huffman2023gpcp}) and semi-hourly data from the Global Precipitation Measurements Integrated Multi-satellitE Retrievals version 7 (GPM IMERG cons / cons2, Tab. S7, \citeA{huffman2023imerg}) to compare our model simulations against observations. 
We note that all coarse model runs, $\overline{\mathrm{DNN}}$-SP-CESM, DNN-ens-SP-CESM and ZM-CESM, are based on one-month spin-up, while SP-CESM is based on a decade-long model run. Due to the short duration of our hybrid simulations below half a year and not imposing Gaussianity of precipitation fields on such short time scales, we focus in the following on the analysis of the precipitation distribution and related percentiles. More common precipitation metrics on climate time scales like mean fields and the monthly means of zonal average precipitation can be found in the supporting information and we will refer to them in the following. Figure \ref{fig:online_precip_distrib} shows zonal averages of the median precipitation (Fig. \ref{fig:online_precip_distrib}a), as well as zonal averages of higher percentiles (Fig. \ref{fig:online_precip_distrib}b,c). To investigate the influence of the internal variability and synoptic features of each simulation on the zonal structures of the respective curves, we add uncertainty ranges based on 50 bootstrapped subsamples of 2000 random time steps ($\sim$ 41 days). Additionally, we show the precipitation probability distribution accumulated across all grid cells and time steps and binned as a function of the baseline precipitation distribution simulated with SP-CESM (Fig. \ref{fig:online_precip_distrib}d). To enable an in-depth evaluation we add the respective percentile curves of GPM IMERG cons and GPCP3.2 daily cons in Figure \ref{fig:online_precip_distrib}a-c, based on a first order conservative remapping on the CESM2 grid. 

$\overline{\mathrm{DNN}}$-SP-CESM and DNN-ens-SP-CESM overestimate the median precipitation in almost the entire extra-tropics compared to all observations and other CESM2 simulations due to the induced drizzle bias from the spurious DNN (Fig. \ref{fig:online_precip_distrib}a). In the tropics the two multi-member parameterizations have a smaller bias of median precipitation than ZM-CESM compared to GPM IMERG cons or SP-CESM, but the main ITCZ related precipitation maximum is displaced to the southern hemisphere (Fig. \ref{fig:online_precip_distrib}a). This southward shift of the first tropical precipitation maximum for $\overline{\mathrm{DNN}}$-SP-CESM and DNN-ens-SP-CESM and a general underestimation of the main ITCZ peak is also visible for higher percentiles compared to GPM IMERG cons (Figs. \ref{fig:online_precip_distrib}b,c;S49).

To better understand these shifts in precipitation patterns seen for $\overline{\mathrm{DNN}}$-SP-CESM and DNN-ens-SP-CESM, we investigate next the mean state of large-scale thermodynamic fields (Figs. S39 - S44), global mean precipitations maps (Fig. S45), the monthly mean evolution of global precipitation patterns (Fig. S46) and global median precipitations maps (Fig. S47). Moreover we evaluate shifts of the extreme precipitation patterns via computing global maps of the 99.9$^{th}$ percentile of precipitation and related zonal averages (Figs. S48,S49). Figures S39 to S41 indicate a pronounced dry bias near the equator and slightly north in the proximity of the main ITCZ peak between 400 hPa over ocean or 700 hPa over land and 950 hPa for $\overline{\mathrm{DNN}}$-SP-CESM and DNN-ens-SP-CESM. Especially over land (Fig. 39), we see an anomalous northward shift of moisture in the lower mid troposphere (500 - 800 hPa) from the equator towards the subtropics for the two multi-member parameterizations. The same is also visible over the ocean and on the southern hemisphere but with weaker magnitude (Figs. S39,S40). This export of moisture reduces the specific humidity near the equator and weakens the amount of precipitable water near the main ITCZ peak at approximately 5$^{\circ}$ N for $\overline{\mathrm{DNN}}$-SP-CESM and DNN-ens-SP-CESM. As a result we see a dampened main ITCZ peak (Figs. S45 - S49) and a weaker migration signal of the ITCZ from the southern hemisphere towards the northern hemisphere (Fig. S46) compared to all other simulations and observations. Despite these biases there is a pronounced reduction of the too strong median precipitation present for ZM-CESM  with $\overline{\mathrm{DNN}}$-SP-CESM and DNN-ens-SP-CESM, over the Maritime continent and tropical southwestern Pacific Ocean near the dateline that is more in agreement with SP-CESM and observations (Fig. S47). This results in lower RMSE of median precipitation globally of $\overline{\mathrm{DNN}}$-SP-CESM and DNN-ens-SP-CESM with respect to GPM IMERG cons or SP-CESM compared to ZM-CESM. Nevertheless Figure S47 clearly shows the drizzle biases of $\overline{\mathrm{DNN}}$-SP-CESM and DNN-ens-SP-CESM due to the spurious DNN in the extra-tropics. Likewise $\overline{\mathrm{DNN}}$-SP-CESM and DNN-ens-SP-CESM underestimate both the mean (Fig. S45), median precipitation (Fig. S47) and extreme precipitation (Fig. S48) over Amazonia and the Congo basin compared to SP-CESM and GPM IMERG cons. This may well be a result of the northward moisture transport towards the subtropics. For the 99.9$^{th}$ percentile we see a similar underestimation of the main ITCZ peak over the tropical Pacific Ocean, the Indian Ocean and the tropical Atlantic Ocean west of equatorial Africa of $\overline{\mathrm{DNN}}$-SP-CESM and DNN-ens-SP-CESM with respect to SP-CESM and GPM IMERG cons (Figs. S48,S49). Moreover we find considerable temperature biases of $\overline{\mathrm{DNN}}$-SP-CESM and DNN-ens-SP-CESM over Antarctica and in the stratosphere of up to 20 K with respect to SP-CESM, that are not present for ZM-CESM (Figs. S42 - S44).

Despite these clear limitations of $\overline{\mathrm{DNN}}$-SP-CESM and DNN-ens-SP-CESM, the two parameterizations enhance the fidelity of extreme precipitation modeling in the tropics compared to ZM-CESM (Figs. \ref{fig:online_precip_distrib}d,S48-S51) and alleviate the known overestimation of intermediate precipitation events ($0.08 ~\frac{mm}{h} ~<~ \mathrm{Prec}~<~ 0.3~\frac{mm}{h}$) of ZM-CESM (Fig. \ref{fig:online_precip_distrib}d). By taking into account GPM IMERG cons we see that a larger similarity to SP-CESM of our schemes helps to reduce extreme precipitation biases in the tropics present in CESM2 (Figs \ref{fig:online_precip_distrib}d, S48-S51) on the one hand. But on the other hand this results in a second precipitation maximum in the tropics at 15$^\circ$ S that is too south and generally too strong with $\overline{\mathrm{DNN}}$-SP-CESM as seen with SP-CESM (Figs. \ref{fig:online_precip_distrib}b,c;S49). This may well be related to the known double ITCZ bias of SP-CESM with respect to observations (Fig. S45, \citeA{Woefle2018}). Along the mid-latitude storm tracks all simulations underestimate extreme precipitation, mostly pronounced over the Southern Ocean at 45$^{\circ}$ S compared to GPM IMERG cons (Figs. S48,S49), while the deviations with respect to GPM IMERG cons are smaller along the northern hemispheric storm track. In contrast to previous findings for the tropics, $\overline{\mathrm{DNN}}$-SP-CESM and DNN-ens-SP-CESM show the lowest reproduction of extreme precipitation in the extra-tropics and the largest biases with respect to GPM IMERG cons (Figs. S48-S51).

\begin{figure}
    \centering
    \includegraphics[width=\textwidth]{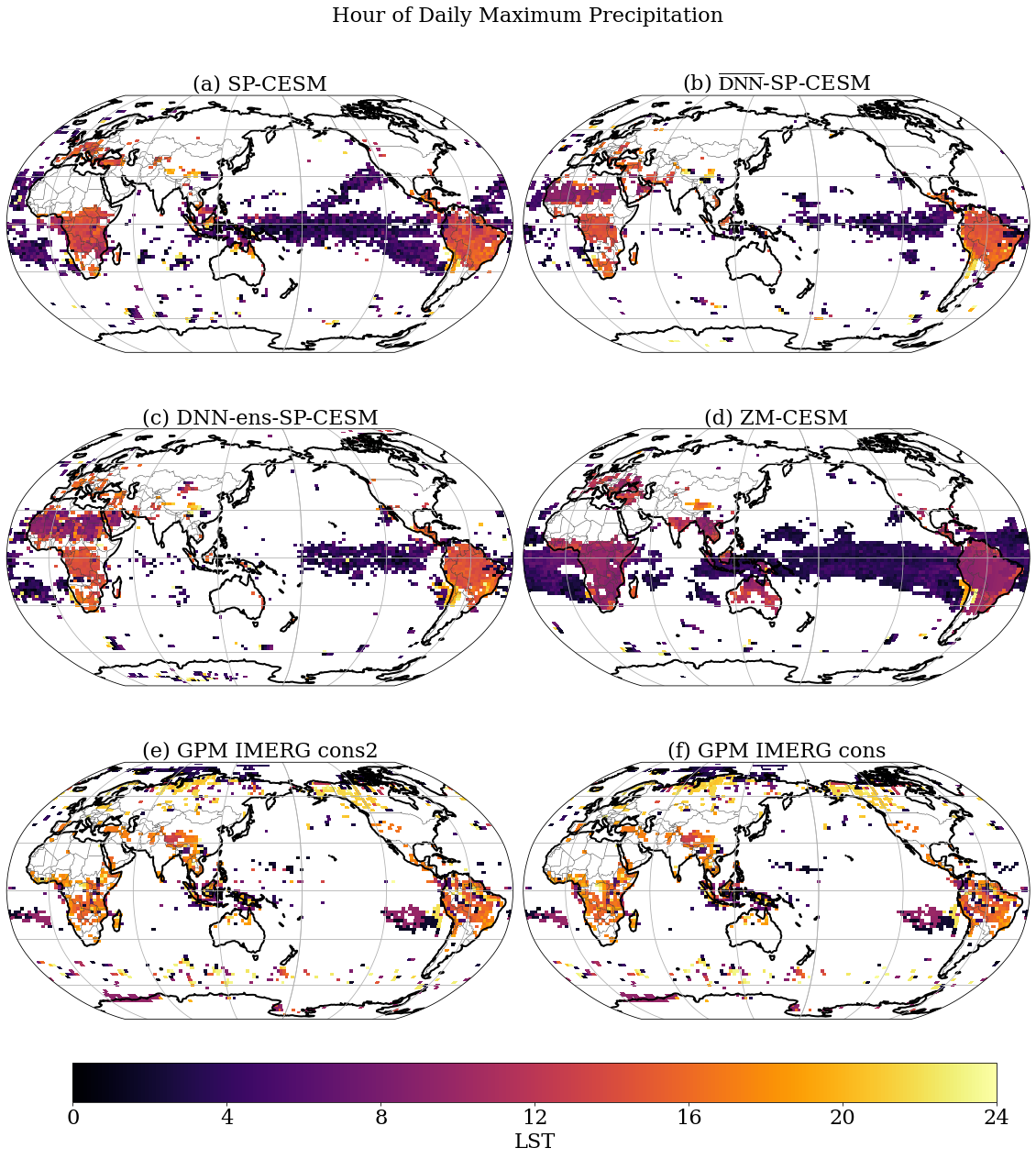}
    \caption{Global maps of the hour of the daily maximum precipitation in the CESM2 runs with the superparametrization SP-CESM (panel a), the deterministic $\overline{\mathrm{DNN}}$ parameterization (panel b), the stochastic DNN-ensemble parameterization (c), the superparametrization SP-CESM (c) and the traditional Zhang-McFarlane scheme (d) and a second order conservative remapping (e) or a first order conservative remapping of GPM IMERG (f) analysed for the period from February to May 2013. The color-coding reveals the diurnal peak in precipitation in local solar time (LST) in areas with a pronounced diurnal cycle of precipitation with a magnitude over a certain threshold, similar to that used in \citeA{Mooers2021}. The white areas in each map show where the magnitude of the diurnal cycle of precipitation is below the threshold.}
    \label{fig:diurnal_cycle}
\end{figure}

As a last step we evaluate the diurnal cycle of all simulations with respect to the GPM IMERG observational product. To consider differences due to the regridding of GPM IMERG, Figure \ref{fig:diurnal_cycle} displays the results of the analysis based on the second order and first order conservative regridding of GPM IMERG to the CESM2 grid. 
Figure \ref{fig:diurnal_cycle} and the regionalized maps over the US, Europe, Amazonia, tropical Africa, India and Southern Africa in Figure S53 show considerable differences of diurnal precipitation between the simulations and GPM IMERG cons or GPM IMERG cons2 for the period February to May 2013. This shows that a diurnal cycle analysis has to take into account the driving seasonality of precipitation patterns and also deviations in the diurnal cycles of simulations and observations. Apart from many differences and biases that Figure \ref{fig:diurnal_cycle} and Figure S53 reveal, GPM IMERG, SP-CESM, $\overline{\mathrm{DNN}}$-SP-CESM and DNN-ens-SP-CESM show a similar afternoon peak of precipitation over Amazonia (Fig. {S53a}), the Congo basin (Fig. {S53b}) and Southern Europe (Fig. {S53c}) for the period February to May 2013. In contrast, ZM-CESM simulates a peak precipitation around noon over these distinct regions. Over the US (Fig. {S53d}) the two multi-member schemes reproduce a similar double peak precipitation diurnal cycle like SP-CESM but with large differences to the observed diurnal cycle of GPCP IMERG independent of the applied regridding. Over Southern Africa we see a similar diurnality of SP-CESM, $\overline{\mathrm{DNN}}$-SP-CESM and DNN-ens-SP-CESM with GPM IMERG (Fig. {S53e}), though there exists a pronounced wet bias of the two multi-member schemes (Fig. \ref{fig:diurnal_cycle},S53e). Over India $\overline{\mathrm{DNN}}$-SP-CESM and DNN-ens-SP-CESM simulate the precipitation in the dry season before the monsoon in large agreement with both SP-CESM and GPM IMERG, while ZM-CESM tends to exhibit monsoon-like conditions with a wet bias (Fig. \ref{fig:diurnal_cycle},S53f). $\overline{\mathrm{DNN}}$-SP-CESM and DNN-ens-SP-CESM show a too strong diurnal cycle over North Africa and the Arabian peninsula (Fig. \ref{fig:diurnal_cycle}), which is related to a wet bias on the order of 0.015 $\frac{mm}{h}$ with respect to SP-CESM and GPM IMERG. One striking deviation of simulations and observations exists over the stratocumuli regions. All traditional simulations appear to have a pronounced diurnal cycle over these regions, while GPM IMERG indicates a negligible diurnal cycle based on the chosen threshold. $\overline{\mathrm{DNN}}$-SP-CESM and DNN-ens-SP-CESM tend to have also a less pronounced diurnal cycle, but the visible pattern in Figure \ref{fig:diurnal_cycle} reveals substantial differences to GPM IMERG.

Finally, we place our findings in context by comparing them against \citeA{Han2023} and \citeA{Kochkov2024}. \citeA{Han2023} used deep convolutional residual neural networks to represent heating and moistening tendencies, as well as cloud liquid and ice water in the Community Atmosphere Model version 5 (CAM5) with real geography. Moreover they successfully coupled one ensemble member to CAM5 and conducted a stable 5-year run with it.  $\overline{\mathrm{DNN}}$-SP-CESM and DNN-ens-SP-CESM show a considerably weaker ITCZ compared to \citeA{Han2023}. This is related to larger biases in the large-scale specific humidity fields in this work, especially in the tropics (Fig. {S39 - S41}), compared to \citeA{Han2023}. 
We note that \citeA{Han2023} sidestepped deep learning surface radiative fluxes (not coupled to the land component), whereas in our study this is explicitly implemented and affects the stability of the hybrid models presented here. 

\citeA{Kochkov2024} presented an approach where they built a neural global circulation model by learning weather and climate related variables based on ERA5 reanalysis as training data. They performed stable neural global circulation model simulations over 40 years in an AMIP-like configuration. Their results suggest that they achieved a considerably smaller RMSE of 850 hPa temperature fields compared to climate models in AMIP configuration. Our multi-member schemes have magnitudes larger temperature biases in the lower troposphere than they reported. We further note that their CRPS loss learning is applicable on our problem statement, but does not allow varying batch sizes or learning rates between members of multi-member parameterizations. Despite this, it has to be noted that the two approaches may not be well comparable due to fact that replacing a parameterization in an existing ESM and designing a new ESM involves different problems and technical tasks.       
In summary, $\overline{\mathrm{DNN}}$-SP-CESM and DNN-ens-SP-CESM have an enhanced stability compared to individual DNNs when partially coupled into CESM2. Furthermore both multi-member parameterizations capture precipitation extremes in the tropics and the underlying diurnal cycle in some regions better than the existing convection scheme -- despite the fact that there are important distortions of the mean state rainfall compared to the original superparameterization and observations related to biases in the large-scale thermodynamic fields.

\section{Conclusion \label{sec:Conclusion}}

In recent years, deterministic deep learning algorithms based on single neural networks have demonstrated capabilities to reproduce key features of subgrid convective processes in climate models \cite{Gentine2018, Rasp2018,Mooers2021,Yuval2021, Wang2022b,Brenowitz2019,Clark2022,Kwa2023,Watt_Meyer_2024,Eyring2024}. However, reproducing the full complexity of convective processes, especially in the planetary boundary layer, remains challenging \cite{Gentine2018,Mooers2021,Behrens2022}. It has been speculated that this lower reproduction skill in the lower troposphere is largely related to the determinism of standard deep learning algorithms, neglecting the stochastic nature of convective processes \cite{Mooers2021,Behrens2022}. In this context, data-driven stochastic or ensemble approaches \cite{Christensen_2024} that are scalable and can robustly overcome these issues could help improve Earth System Models (ESMs).

This study presents and evaluates novel deep learning approaches to account for subgrid variability, due to stochasticity, to improve ESMs. We demonstrate that the uncertainty and variability of such processes, as represented by the Superparameterized Community Earth System Model 2 (SPCESM2), can be captured via multi-member parameterizations combining predictions using Deep Neural Networks (DNNs) or Variational Encoder Decoders (VEDs). This variability in unresolved convective processes is particularly relevant in the lower troposphere associated with turbulence and shallow convection, as well as in the upper troposphere and lower stratosphere due to deep convection. We focus particularly our analysis on these challenging levels, where multi-member and stochastic parameterizations exhibit a pronounced overconfidence or underdispersion of capturing the full variability related to convection. There is, however, a trade-off between capturing the uncertainty of subgrid processes and their mean effect on the system, affecting the overall performance of the deep learned parameterization. A DNN with active Monte Carlo dropout during training and prediction neither fully captures the variability of unresolved processes nor it is as accurate as other deep learning algorithms explored here. Perturbing the latent space of VEDs provides a good uncertainty range in their predictions, though accuracy in their predictions is substantially affected. Randomly drawing a subset of predictions from different DNNs, DNN-ensemble (Tab. \ref{tab:overview_stoch_deter}), enables us to achieve both a calibrated uncertainty compared to the superparameterized ESM and skillful predictions as good as using the full deterministic multi-member parameterization, $\overline{\mathrm{DNN}}$ (Tab. \ref{tab:overview_stoch_deter}).

We, therefore, couple the best performing stochastic deep learned parameterization, DNN-ensemble, as well as its deterministic counterpart, $\mathrm{\overline{DNN}}$, to the coarse ESM host model. Our hybrid simulations are designed as preliminary experiments toward developing a stable multi-member parameterization of subgrid processes in an ESM. While the proposed approach can be further improved and warrants future work, our study identifies several key challenges that should be addressed moving forward. First, the coupling of the entire set of output variables $\boldsymbol{Y^{pred}}$ remains challenging. The related hybrid runs with the deep learned multi-member parameterizations are stable only over a few days - that illustrates the need for ``perfect condensate'' experiments. In these experiments we partially coupled our parameterizations including key surface radiative fluxes for surface coupling while condensate tendencies are simulated with the superparameterization running alongside. Secondly, we acknowledge that these partially coupled multi-member parameterizations are clearly more computational expensive than traditional parameterizations and slower than the superparameterization itself. With this pragmatic though computationally demanding approach we conduct hybrid simulations for a duration of approximately five months with DNN-ensemble and $\mathrm{\overline{DNN}}$. In contrast, simulations with individual DNNs fail within the first five days in most cases. To benchmark our simulations we use the precipitation observations of GPM IMERG version 7 (GPM IMERG cons, \citeA{huffman2023imerg}) and GPCP version 3.2 (GPCP3.2 daily cons, \citeA{huffman2023gpcp}) or the superparameterization (SP-CESM) as ground truth and the traditional Zhang-McFarlane scheme (ZM-CESM, \citeA{ZhangMcFarlane1995}).

Our multi-member parameterizations capture large-scale thermodynamic patterns but exhibit temperature biases at the surface over Antarctica and in the stratosphere (exceeding +20 K), as well as a negative specific humidity bias near the equator in the troposphere. The latter is associated with an anomalous northward moisture transport from the equator over land in the lower free troposphere. These two latter deficiencies contribute to a general weakening of the primary peak of the ITCZ and a dampening of its seasonal migration from the southern to the northern hemisphere seen in precipitation fields. Our multi-member parameterizations underestimate precipitation over tropical continents and extreme precipitation in the extra-tropics compared to SP-CESM and GPM IMERG cons. Despite these challenges, hybrid simulations with $\mathrm{\overline{DNN}}$ and DNN-ensemble reduce some biases of ZM-CESM with respect to the underestimation of precipitation extremes and overestimation of intermediate precipitation in the tropics. This result is potentially influenced by the reliance on predicting condensate tendency with the superparameterization in our hybrid simulations. Moreover, our multi-member parameterizations show some improvements compared to ZM-CESM over (tropical) continents with respect to the diurnal cycle of precipitation, e.g. shifting the too early peaks towards the afternoon. Nevertheless, we see large discrepancies between the simulations including SP-CESM and GPM IMERG cons with respect to the governing diurnal cycle in the period February to May, which makes the interpretability of the precipitation diurnal cycle of simulations without observations challenging.    


The results of our study indicate that large room for improvements and several open questions remain. However, there exist potential ways forward to alleviate some limitations of our hybrid simulations based on recent advances in machine learning subgrid atmospheric processes. 
First of all, in an ideal case an operational hybrid model, with deterministic or stochastic deep learning parameterizations, would run stably without blowups or climate drifts (systematic and increasing long-term errors). \citeA{Han2023}, \citeA{Wang2022}, \citeA{Kochkov2024}, \citeA{Yu2024} and \citeA{Hu2024} proved that this is possible with realistic boundary conditions over several years. However, while \citeA{Wang2022} used an atmosphere only configuration, \citeA{Han2023} neglected radiative fluxes important for atmosphere-land coupling. \citeA{Kochkov2024} showed that stability with small temperature biases over decades in an AMIP-like configuration is achievable, and explored learning atmospheric processes in a stochastic framework that exceeds the framework presented in our manuscript. \citeA{Yuval2021} found that separating the emulation of condensate tendencies from the emulation of surface variables in independent neural networks yields the successful emulation of the first and stable hybrid simulations in an aquaplanet configuration. Recently, \citeA{Yu2024} and \citeA{Hu2024} showed that learning condensate tendencies and successfully coupling these with an ESM without a multi-member parameterizations is possible in a similar SP setup, but requires more advanced losses and deep learning algorithms. Future work will aim to further develop deep learning parameterizations and build up on existing deep learning schemes, including the stochastic approaches proposed here, to enable accurate long-term \textit{hybrid model} simulations. Another open question of our manuscript is how to increase the reproduction skill of cloud water and cloud ice water tendencies with deep learning models. Potential approaches may include: replacing DNNs and other rather simplistic models with more complex U-Net architectures combined with fine tuned loss functions \cite{Yu2024,Hu2024}, splitting the emulation task into smaller sub-tasks with dedicated models \cite{Yuval2021}, substituting deterministic metrics in the loss function for proper scoring metrics such as the Continuous Rank Probability Score \cite{Kochkov2024}, using loss functions that maximize likelihood \cite{Haynes2023}, or applying novel probabilistic data-driven models \cite{Christensen_2024}. A community benchmark dataset has been released that should facilitate intercomparisons between future advances in machine learning parameterizations for ESMs with state-of-the-art algorithms \cite{Yu2023_neurips}. Likewise, the use of a more flexible Fortran-Python coupler as presented in \citeA{Yu2024} and \citeA{Hu2024} may enable us to explore the potential of latent space perturbation with VEDs to obtain more calibrated uncertainty quantification of convective processes also in coupled simulations. Looking ahead, our multi-member parameterization could be used to assess confidence in its own predictions during hybrid simulations (see Figs.~S54,S55). This approach is similar to that of \citeA{Mansfield2024}, who demonstrated that a coupled neural network can assess quasi-biennial oscillation uncertainty online in the context of gravity wave parameterization, and to \citeA{sanford2023}, who trained a one-class support vector machine to flag anomalous predictions on-the-fly that do not warrant machine learning correction of atmospheric profiles.

This work demonstrates that hybrid simulations of deterministic and stochastic deep leaning multi-member parameterizations with a complete coupling of subgrid radiative fluxes to a comprehensive land model are stable over a period of five months. The provided issues of emulating condensate tendencies are sidestepped but our computationally demanding online experiments compared to a traditional convection parameterization show large biases in thermodynamic state variables that need to be addressed in future work. We show that deep learning multi-member parameterizations improve the representation of convective processes based on test data, especially within the planetary boundary layer, compared to individual neural networks. We further demonstrate that this translates into an enhanced online stability of multi-member parameterizations compared to individual networks that are members of the parameterization. In ESM simulations we find that our multi-member parameterizations introduce biases that result in an unrealistic ITCZ compared to observations. However we see 
improvements with our multi-member schemes in the reproduction of precipitation extremes in the tropics and the diurnal cycle of precipitation over (tropical) continents compared to a traditional convection scheme. Multi-member and other developed stochastic parameterizations further have the potential to add to each prediction and variable a related uncertainty quantification. These are key steps forward to increase the quality of simulated complex processes like convection and the trustworthiness of deep learning parameterizations in general that will be developed for the next generation of Earth System Models.

\appendix
\section{Benchmarking and Coupling strategy}
\label{Appendix_A}
\subsection{Benchmarking}
\label{sec:Ens_metr}

We evaluate the quality of the spread given by the different stochastic and deterministic parameterizations via uncertainty quantification with respect to the test data using three metrics. Specifically, we quantify the \textit{aleatoric uncertainty} associated with the randomness aspect of the data-generation process, including the chaotic nature of convective processes in the atmosphere \cite{Haynes2023}. Firstly, we use the Continuous Rank Probability Score (CRPS), which is the difference between the MAE (first term) and the spread inside the ensemble (second term) in equation \ref{eq:CRPS_def} \cite{Haynes2023}:

\begin{linenomath*}
    \begin{equation}\label{eq:CRPS_def}
        \mathrm{CRPS}={\frac{1}{\mathrm{N}_{ens}}}\sum_{i=1}^{\mathrm{N}_{ens}} |\boldsymbol{Y}^{pred}_i-\boldsymbol{Y}|-{\frac{1}{2\mathrm{N}_{ens}^2}}\sum_{i=1}^{\mathrm{N}_{ens}}\sum_{l=1}^{\mathrm{N}_{ens}}\ |\boldsymbol{Y}^{pred}_i-\boldsymbol{Y}^{pred}_l|
    \end{equation}
\end{linenomath*}

CRPS is both sensitive to the deterministic quality of individual predictions $\boldsymbol{Y}_i^{pred}$ condensed in the MAE term and to the spread of the predictions inside the ensemble. This makes CRPS a suitable stochastic loss function for deep learning \cite{Haynes2023,Kochkov2024}. However for our parameterization task CRPS is not perfectly suited, as we have to use uniform batch sizes and learning rates during training across multi-members to apply CRPS. Moreover CRPS in its probabilistic integral form is a proper score \cite{Gneiting2007} of negative orientation with a fixed lower bound of 0 (perfect skill) and upper bound 1 (no skill) for classification tasks. The analytic version of CRPS used in regression tasks, shown in equation \ref{eq:CRPS_def}, has in contrast only a fixed lower bound of 0 (perfect skill) but not necessarily a finite upper bound.

Secondly, we use spread-skill diagrams to evaluate whether the skill of the stochastic and deterministic ensemble predictions (measured with the Root Mean Squared Error, RMSE) is correlated with the ensemble spread \cite{Haynes2023}. An ideal ensemble would have a pronounced correlation between spread and RMSE with a spread-skill ratio of one \cite{Haynes2023,Berner2017}. To sort the magnitude of the spread of the parameterizations for given $\boldsymbol{X}$ and selected output variables of interest $Y^{pred}_{i,j}$,  we bin the spread into a number of classes $\mathrm{N}_{\mathrm{bins}}$ and compute the bin average for each class (equation \ref{eq:Spread_def}). Then we calculate the conditionally averaged RMSE (equation \ref{eq:RMSE_def}) for each class $b \in \llbracket 1,\mathrm{N}_{\mathrm{bins}}\rrbracket$:

\begin{linenomath*}
    \begin{equation}\label{eq:Spread_def}
        \mathrm{Spread}_{j,b}={\frac{1}{\mathrm{N}_{\mathrm{counts},b}}}\sum_{c=1}^{\mathrm{N}_{\mathrm{counts},b}} \sqrt{\frac{1}{\mathrm{N}_{ens}-1}\sum_{i=1}^{\mathrm{N}_{ens}}\left(\overline{Y^{pred}_{c,j}}-Y^{pred}_{c,i,j}\right)^2}
    \end{equation}
\end{linenomath*}

\begin{linenomath*}
    \begin{equation}\label{eq:RMSE_def}
        \mathrm{RMSE}_{j,b}=\sqrt{{\frac{1}{\mathrm{N}_{\mathrm{counts},b}}}\sum_{c=1}^{\mathrm{N}_{\mathrm{counts},b}} \left(Y_{c,j}-\overline{Y^{pred}_{c,j}}\right)^2},
    \end{equation}
\end{linenomath*}
where $\mathrm{N}_{\mathrm{bins}}$ is the number of classes (bins) and $\mathrm{N}_{\mathrm{counts},b}$ represents the number of elements within a class $b \in \llbracket 1,\mathrm{N}_{\mathrm{bins}}\rrbracket$.

Finally, we calculate the probability integral transform (PIT). This metric is similar to rank histograms, where the true value $Y_{j}$ is ranked within the ensemble $\{Y^{pred}_j\}$ (i.e., the test data sample is situated between the (r-1)$^{th}$ and r$^{th}$ ensemble member and gets the rank r, where r is the rank ID). The PIT diagram is then obtained by computing the probability density function of all observed ranks $r \in \llbracket 1, N_{ens} \rrbracket$ of $Y_{j}$ (a probability value of each rank r; the y-axis) binned by the PIT values of each rank r (defined by the CDF of all ranks $N_{ens}$, x-axis). We use the PIT to evaluate whether the ensemble is ``overdispersive'' (underconfident, meaning that the ensemble overestimates the variations in $Y_{j}$) 
or ``underdispersive'' (overconfident, where the ensemble underestimates the variations in $Y_{j}$). 
Ideally, the PIT curve is a horizontal line with an associated probability of $\frac{1}{\mathrm{N}_{ens}}$, which can be used to compute the PIT distance metric between the actual and ideal PIT case similar to the one shown in \citeA{Haynes2023}.

\subsection{Coupling DNNs to CESM2}
\label{sec:coupling}
To couple our multi-member and stochastic parameterizations into CESM2 (replacing the SP component) we use the Fortran-Keras-Bridge (FKB) \cite{Ott2020}. To enforce the positivity of precipitation and radiative fluxes as predictants, we add a ``positivity layer'' as a constraint layer \cite{Beucler2021} to all DL models of the parameterizations. By design our DL models could predict negative values due to the used loss functions and related epistemic uncertainty (e.g. that the DL models predict in out-of-sample conditions). The ``positivity layer'' maps these variables with a ReLU activation to positive values. This ensures that the predicted precipitation and radiative fluxes of our parameterizations are positive finite and enable a successful integration in CESM2.  
We restrict our online coupling efforts to the deterministic and stochastic multi-member DNN parameterizations, which show superior offline performance compared to other developed parameterizations in section \ref{sec:Offline_Results}. First we transform the native weights and biases files into text files, which makes the files accessible for FKB and related Fortran compilers \cite{Ott2020}. Then we create a standalone repository that allows to couple individual DNNs, $\overline{\mathrm{DNN}}$ and DNN-ensemble into CESM2. For DNN-ensemble we implement a random average function on the grid cell level. In initial coupled experiments we find in some cases unrealistic simulated solar and shortwave radiative fluxes of more than 50 $\frac{W}{m^2}$ during night-time conditions on lower latitudes. We relate these deficiencies to one particular DNN that struggled with the test set and shows also poor qualitative online performance. Surprisingly this DNN has by far the most vigorous online stability of all DNNs ({Figs. S37,S38}). To enhance the robustness of the online runs and the interpretability of simulated processes we enforce realistic radiative conditions for the coupling to the land and ocean surface by setting all solar fluxes and shortwave fluxes included in $ \boldsymbol{Y_{\mathrm{rad}}}$ (equation \ref{eq:Y_rad}) to zero $\frac{W}{m^2}$, if the cosine of the zenith angle of the incoming solar radiation in CESM2 at the current time step and grid cell is zero or negative (night-time conditions). Additionally, we implement a partial coupling scheme of our parameterization for certain variables, while other variables are simulated with the SP running aside. Our best performing setup that we present in section \ref{sec:online} relies on coupling all predicted variables from our parameterizations into CESM2 except for cloud ice water $\boldsymbol{\dot{q}_{ci}}$ and cloud liquid water tendency $\boldsymbol{\dot{q}_{cl}}$, which remain simulated by SP. This partial coupling is computationally demanding (Tab. S6), but stabilizes online simulations, e.g., increasing the time until CESM2 crashes with our parameterization from the order of days or hours to more than five months (see section \ref{sec:online}). For the online runs we use the predefined time stepping of SPCESM, with a native CESM2 time step of 1800 s and an SP time step of 20 s. The subgrid source terms coming from SP and our parameterization are updated at every CESM2 time step. We perform CESM2 simulations based on initialisation files of January 2013 that included one month of SP spin-up, which is necessary for a realistic representation of global precipitation patterns. Our simulations start at the beginning of February. This coincides with the conditions that individual DNNs are optimized for during the training, as the respective data set contains the first seven days of each month of the year 2013. Nevertheless we tested also additional initialisation dates and found that the stability of our multi-member parameterizations is sensitive to the choice of the initialisation date and our multi-member parameterizations cause initialisation shocks for other dates. 

%
%

%

%

\section{Open Research}

The code used to train all DNNs, all VEDs, build the multi-member and stochastic parameterizations and to produce all figures of this manuscript is accessible via a Github repository, which is archived with Zenodo \cite{behrens_2024_github_repo}. The repository includes the Jupyter Notebooks, python files, run scripts for the online simulations of the machine learning parameterizations, conda environments used to reproduce all figures of the manuscript and attached supporting information. The text file $List\_of\_Figures.txt$ illustrates where to find the code to reproduce each figure in the Github repository. 
The above mentioned Github repository is based on Stephan Rasp's repository published for \citeA{Rasp2018}, which can be found on Github and is archived using Zenodo \cite{stephan_rasp_2018_1402384}. The repository includes a helpful quickstart guide (\textit{quickstart.ipynb}) to preprocess raw SPCAM data that is similar to SPCESM2 data, train a neural network and to show first steps how to evaluate the neural network.

An example of SPCESM2 data was archived on Zenodo for this publication \cite{behrens_2024_SPCESM_data}. This folder includes raw SPCESM2 data, preprocessed data and initialization files produced for this publication. The full SPESM2 raw and preprocessed data, of the order of several TBs, is archived on DKRZ and available upon request. The data of the conducted hybrid simulations is also archived on DKRZ and available upon request. 

\acknowledgments
We thank Dr. Axel Lauer and three anonymous reviewers for their helpful comments and suggestions, which improved our manuscript.

GB, FIS, PG, MS and VE received funding for this study from the European Research Council (ERC) Synergy Grant ``Understanding and modeling the Earth System with Machine Learning (USMILE)'' under the Horizon 2020 research and innovation programme (Grant agreement No. 855187).
GB’s research for this study was also funded by the Deutsche Forschungsgemeinschaft (DFG, German Research Foundation) through the Gottfried Wilhelm Leibniz Prize awarded to Veronika Eyring (Reference No. EY 22/2-1).
TB acknowledges partial funding from the Swiss State Secretariat for Education, Research and Innovation (SERI) for the Horizon Europe project AI4PEX (Grant agreement ID: 101137682).
This publication is part of the EERIE project (Grant Agreement No 101081383) funded by the European Union. Views and opinions expressed are however those of the author(s) only and do not necessarily reflect those of the European Union or the European Climate Infrastructure and Environment Executive Agency (CINEA). Neither the European Union nor the granting authority can be held responsible for them. 
This work has received funding from the SERI under contract \#22.00366. 
This work was funded by UK Research and Innovation (UKRI) under the UK government’s Horizon Europe funding guarantee (grant number 10057890, 10049639, 10040510, 10040984).

This work used resources of the Deutsches Klimarechenzentrum (DKRZ) granted by its Scientific Steering Committee (WLA) under project ID 1179 (USMILE) and 1083 (Climate Informatics), and the supercomputer JUWELS at
the Jülich Supercomputing Centre (JSC) under the Earth System Modelling Project (ESM).

PG, MP, and SY acknowledge funding from  National Science Foundation (NSF) Science and Technology Center (STC) Learning the Earth with Artificial Intelligence and Physics (LEAP), Award \# 2019625-STC. MP and SY were additionally supported by fundings from U.S. Department of Energy Grants (the EAGLES project 74358, the Exascale Computing Project 17-SC-20-SC, DE-SC0023368, and DE-SC0022331).


%
%



\bibliography{main_behrens_2023.bbl}

%
%
%
%
%


\includegraphics[scale=0.85,page=1]{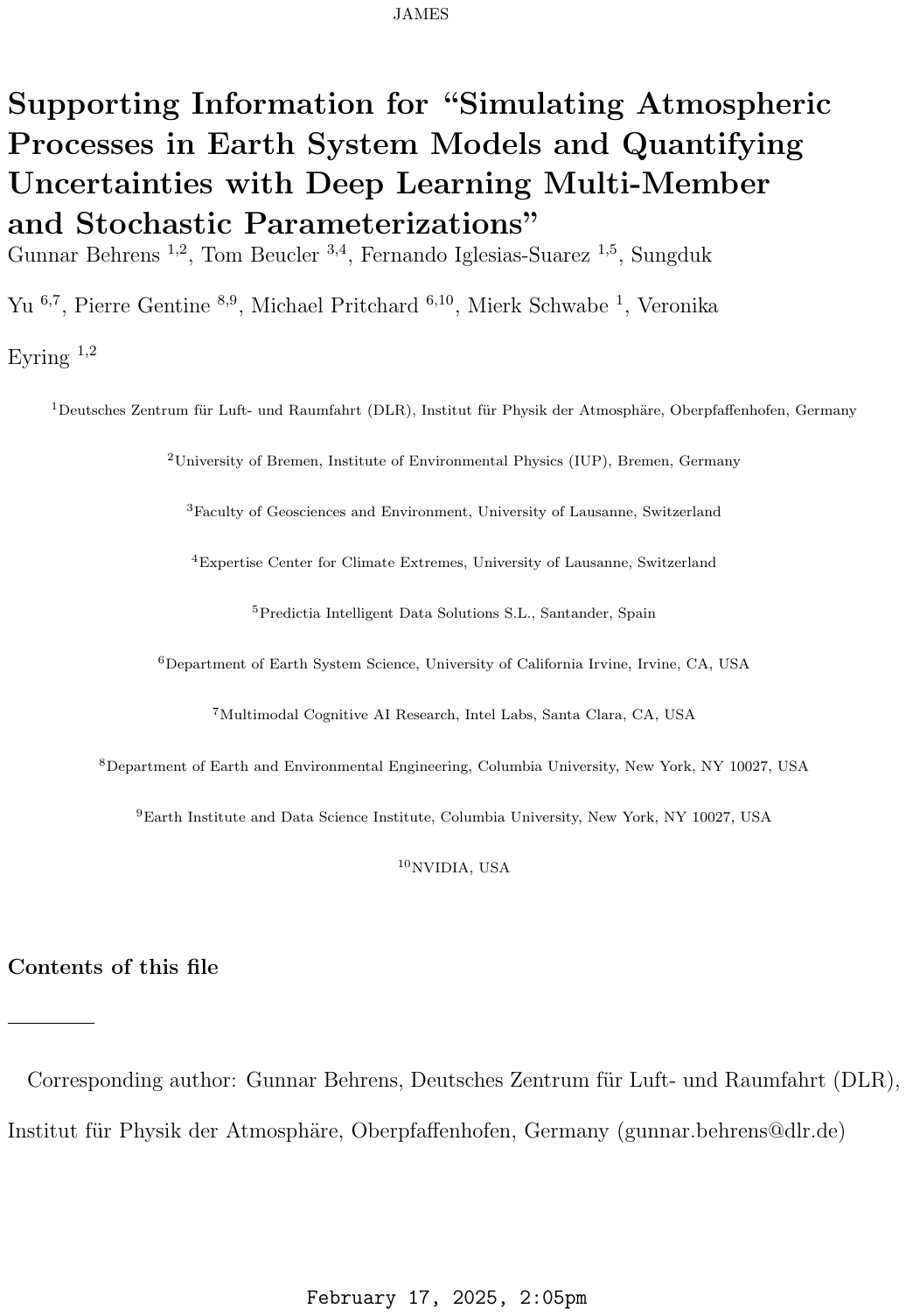}
\includegraphics[scale=0.85,page=2]{SI_Behrens25_stochastic_parametrization_for_SPCESM.pdf}
\includegraphics[scale=0.85,page=1]{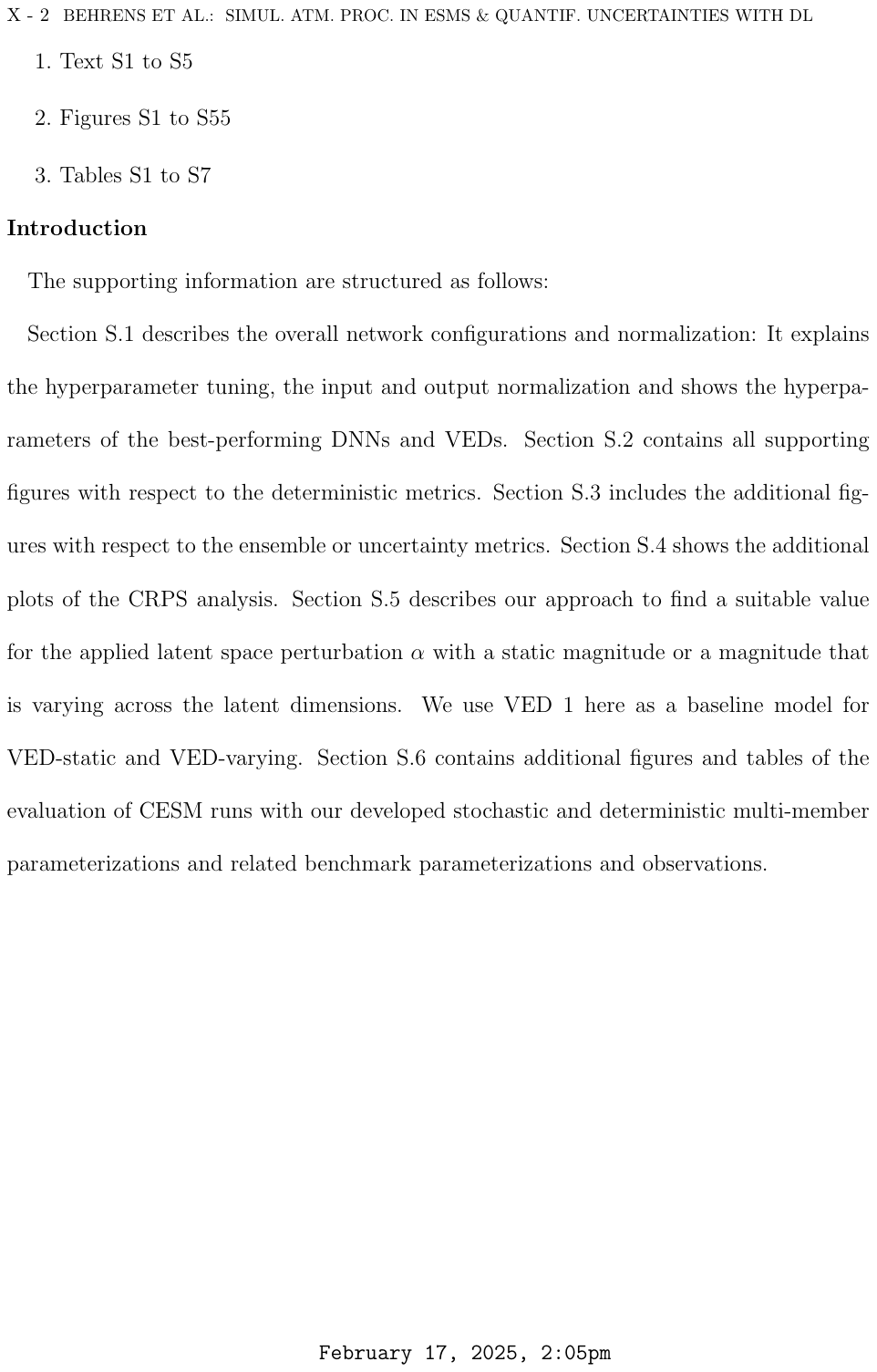}
\includegraphics[scale=0.85,page=3]{SI_Behrens25_stochastic_parametrization_for_SPCESM.pdf}
\includegraphics[scale=0.85,page=4]{SI_Behrens25_stochastic_parametrization_for_SPCESM.pdf}
\includegraphics[scale=0.85,page=5]{SI_Behrens25_stochastic_parametrization_for_SPCESM.pdf}
\includegraphics[scale=0.85,page=6]{SI_Behrens25_stochastic_parametrization_for_SPCESM.pdf}
\includegraphics[scale=0.85,page=7]{SI_Behrens25_stochastic_parametrization_for_SPCESM.pdf}
\includegraphics[scale=0.85,page=8]{SI_Behrens25_stochastic_parametrization_for_SPCESM.pdf}
\includegraphics[scale=0.85,page=9]{SI_Behrens25_stochastic_parametrization_for_SPCESM.pdf}
\includegraphics[scale=0.85,page=10]{SI_Behrens25_stochastic_parametrization_for_SPCESM.pdf}
\includegraphics[scale=0.85,page=11]{SI_Behrens25_stochastic_parametrization_for_SPCESM.pdf}
\includegraphics[scale=0.85,page=12]{SI_Behrens25_stochastic_parametrization_for_SPCESM.pdf}
\includegraphics[scale=0.85,page=13]{SI_Behrens25_stochastic_parametrization_for_SPCESM.pdf}
\includegraphics[scale=0.85,page=14]{SI_Behrens25_stochastic_parametrization_for_SPCESM.pdf}
\includegraphics[scale=0.85,page=15]{SI_Behrens25_stochastic_parametrization_for_SPCESM.pdf}
\includegraphics[scale=0.85,page=16]{SI_Behrens25_stochastic_parametrization_for_SPCESM.pdf}
\includegraphics[scale=0.85,page=17]{SI_Behrens25_stochastic_parametrization_for_SPCESM.pdf}
\includegraphics[scale=0.85,page=18]{SI_Behrens25_stochastic_parametrization_for_SPCESM.pdf}
\includegraphics[scale=0.85,page=19]{SI_Behrens25_stochastic_parametrization_for_SPCESM.pdf}
\includegraphics[scale=0.85,page=20]{SI_Behrens25_stochastic_parametrization_for_SPCESM.pdf}
\includegraphics[scale=0.85,page=21]{SI_Behrens25_stochastic_parametrization_for_SPCESM.pdf}
\includegraphics[scale=0.85,page=22]{SI_Behrens25_stochastic_parametrization_for_SPCESM.pdf}
\includegraphics[scale=0.85,page=23]{SI_Behrens25_stochastic_parametrization_for_SPCESM.pdf}
\includegraphics[scale=0.85,page=24]{SI_Behrens25_stochastic_parametrization_for_SPCESM.pdf}
\includegraphics[scale=0.85,page=25]{SI_Behrens25_stochastic_parametrization_for_SPCESM.pdf}
\includegraphics[scale=0.85,page=26]{SI_Behrens25_stochastic_parametrization_for_SPCESM.pdf}
\includegraphics[scale=0.85,page=27]{SI_Behrens25_stochastic_parametrization_for_SPCESM.pdf}
\includegraphics[scale=0.85,page=28]{SI_Behrens25_stochastic_parametrization_for_SPCESM.pdf}
\includegraphics[scale=0.85,page=29]{SI_Behrens25_stochastic_parametrization_for_SPCESM.pdf}
\includegraphics[scale=0.85,page=30]{SI_Behrens25_stochastic_parametrization_for_SPCESM.pdf}
\includegraphics[scale=0.85,page=31]{SI_Behrens25_stochastic_parametrization_for_SPCESM.pdf}
\includegraphics[scale=0.85,page=32]{SI_Behrens25_stochastic_parametrization_for_SPCESM.pdf}
\includegraphics[scale=0.85,page=33]{SI_Behrens25_stochastic_parametrization_for_SPCESM.pdf}
\includegraphics[scale=0.85,page=34]{SI_Behrens25_stochastic_parametrization_for_SPCESM.pdf}
\includegraphics[scale=0.85,page=35]{SI_Behrens25_stochastic_parametrization_for_SPCESM.pdf}
\includegraphics[scale=0.85,page=36]{SI_Behrens25_stochastic_parametrization_for_SPCESM.pdf}
\includegraphics[scale=0.85,page=37]{SI_Behrens25_stochastic_parametrization_for_SPCESM.pdf}
\includegraphics[scale=0.85,page=38]{SI_Behrens25_stochastic_parametrization_for_SPCESM.pdf}
\includegraphics[scale=0.85,page=39]{SI_Behrens25_stochastic_parametrization_for_SPCESM.pdf}
\includegraphics[scale=0.85,page=40]{SI_Behrens25_stochastic_parametrization_for_SPCESM.pdf}
\includegraphics[scale=0.85,page=41]{SI_Behrens25_stochastic_parametrization_for_SPCESM.pdf}
\includegraphics[scale=0.85,page=42]{SI_Behrens25_stochastic_parametrization_for_SPCESM.pdf}
\includegraphics[scale=0.85,page=43]{SI_Behrens25_stochastic_parametrization_for_SPCESM.pdf}
\includegraphics[scale=0.85,page=44]{SI_Behrens25_stochastic_parametrization_for_SPCESM.pdf}
\includegraphics[scale=0.85,page=45]{SI_Behrens25_stochastic_parametrization_for_SPCESM.pdf}
\includegraphics[scale=0.85,page=46]{SI_Behrens25_stochastic_parametrization_for_SPCESM.pdf}
\includegraphics[scale=0.85,page=47]{SI_Behrens25_stochastic_parametrization_for_SPCESM.pdf}
\includegraphics[scale=0.85,page=48]{SI_Behrens25_stochastic_parametrization_for_SPCESM.pdf}
\includegraphics[scale=0.85,page=49]{SI_Behrens25_stochastic_parametrization_for_SPCESM.pdf}
\includegraphics[scale=0.85,page=50]{SI_Behrens25_stochastic_parametrization_for_SPCESM.pdf}
\includegraphics[scale=0.85,page=51]{SI_Behrens25_stochastic_parametrization_for_SPCESM.pdf}
\includegraphics[scale=0.85,page=52]{SI_Behrens25_stochastic_parametrization_for_SPCESM.pdf}
\includegraphics[scale=0.85,page=53]{SI_Behrens25_stochastic_parametrization_for_SPCESM.pdf}
\includegraphics[scale=0.85,page=54]{SI_Behrens25_stochastic_parametrization_for_SPCESM.pdf}
\includegraphics[scale=0.85,page=55]{SI_Behrens25_stochastic_parametrization_for_SPCESM.pdf}
\includegraphics[scale=0.85,page=56]{SI_Behrens25_stochastic_parametrization_for_SPCESM.pdf}
\includegraphics[scale=0.85,page=57]{SI_Behrens25_stochastic_parametrization_for_SPCESM.pdf}
\includegraphics[scale=0.85,page=58]{SI_Behrens25_stochastic_parametrization_for_SPCESM.pdf}
\includegraphics[scale=0.85,page=59]{SI_Behrens25_stochastic_parametrization_for_SPCESM.pdf}
\includegraphics[scale=0.85,page=60]{SI_Behrens25_stochastic_parametrization_for_SPCESM.pdf}
\includegraphics[scale=0.85,page=61]{SI_Behrens25_stochastic_parametrization_for_SPCESM.pdf}
\includegraphics[scale=0.85,page=62]{SI_Behrens25_stochastic_parametrization_for_SPCESM.pdf}
\includegraphics[scale=0.85,page=63]{SI_Behrens25_stochastic_parametrization_for_SPCESM.pdf}
\includegraphics[scale=0.85,page=64]{SI_Behrens25_stochastic_parametrization_for_SPCESM.pdf}
\includegraphics[scale=0.85,page=65]{SI_Behrens25_stochastic_parametrization_for_SPCESM.pdf}
\includegraphics[scale=0.85,page=66]{SI_Behrens25_stochastic_parametrization_for_SPCESM.pdf}
\includegraphics[scale=0.85,page=67]{SI_Behrens25_stochastic_parametrization_for_SPCESM.pdf}
\includegraphics[scale=0.85,page=68]{SI_Behrens25_stochastic_parametrization_for_SPCESM.pdf}
\includegraphics[scale=0.85,page=69]{SI_Behrens25_stochastic_parametrization_for_SPCESM.pdf}
\includegraphics[scale=0.85,page=70]{SI_Behrens25_stochastic_parametrization_for_SPCESM.pdf}
\includegraphics[scale=0.85,page=71]{SI_Behrens25_stochastic_parametrization_for_SPCESM.pdf}
\includegraphics[scale=0.85,page=72]{SI_Behrens25_stochastic_parametrization_for_SPCESM.pdf}
\includegraphics[scale=0.85,page=73]{SI_Behrens25_stochastic_parametrization_for_SPCESM.pdf}
\includegraphics[scale=0.85,page=74]{SI_Behrens25_stochastic_parametrization_for_SPCESM.pdf}
\includegraphics[scale=0.85,page=75]{SI_Behrens25_stochastic_parametrization_for_SPCESM.pdf}
\includegraphics[scale=0.85,page=76]{SI_Behrens25_stochastic_parametrization_for_SPCESM.pdf}
\includegraphics[scale=0.85,page=1]{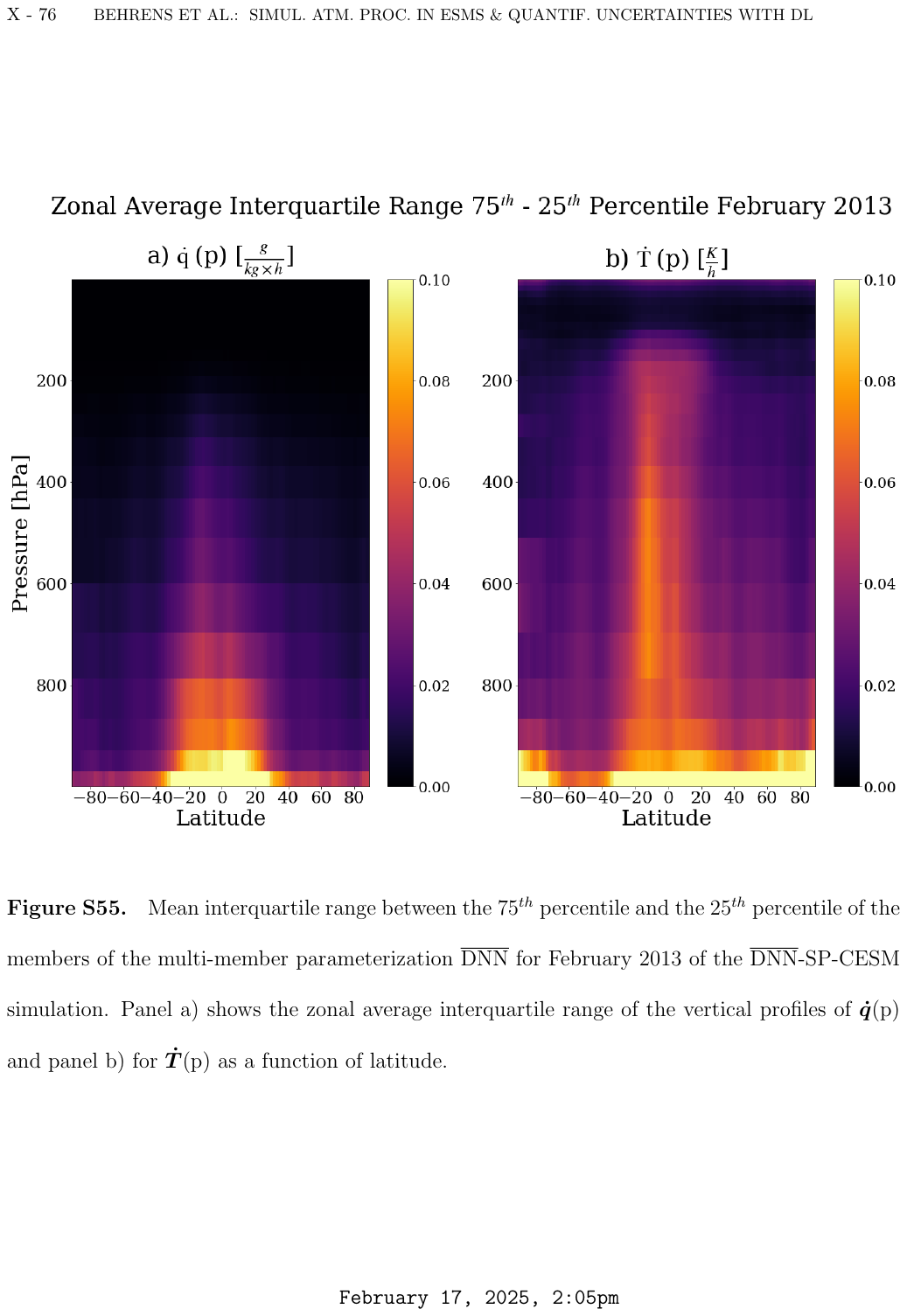}

%
%
%
%
%

\end{document}